\documentstyle[preprint,aps,tighten,epsfig,axodraw]{revtex} %

\begin{document}

\draft

\preprint{\begin{tabular}{l}
\hbox to\hsize{\mbox{ }\hfill hep--ph/9907474}  \\ 
\hbox to\hsize{\mbox{ }\hfill KIAS-P99057}  \\ 
\hbox to\hsize{\mbox{ }\hfill MADPH--99-1130}  \\ 
\hbox to\hsize{\mbox{ }\hfill SNUTP 99-035} \\ 
\hbox to\hsize{\mbox{ }\hfill \today}       \\ 
\hbox to\hsize{\mbox{ }\hfill       }       \\ 
          \end{tabular}}
\vskip 2.5cm

\title{CP Phases in Correlated Production and Decay of Neutralinos\\
       in the Minimal Supersymmetric Standard Model} 

\author{S.Y.~Choi}
\address{Korea Institute for Advanced Study, 207--43, Cheongryangri--dong,
         Dongdaemun--gu, Seoul 130--012, Korea}

\author{H.S.~Song and W.Y.~Song}
\address{Center for Theoretical Physics and Department of Physics,
         Seoul National University, Seoul 151-742, Korea}

\maketitle

\begin{abstract}
We investigate the associated production of neutralinos  
$e^+e^-\rightarrow\tilde{\chi}^0_1\tilde{\chi}^0_2$ accompanied by the
neutralino leptonic decay $\tilde{\chi}^0_2\rightarrow\tilde{\chi}^0_1
\ell^+\ell^-$, taking into account initial beam polarization 
and production-decay spin correlations in the minimal supersymmetric
standard model with general CP phases but without generational mixing 
in the slepton sector. The stringent constraints from the electron EDM
on the CP phases are also included in the discussion. 
Initial beam polarizations lead to three CP--even distributions and
one CP--odd distribution, which can be studied independently of 
the details of the neutralino decays. We find that 
the production cross section and the branching fractions of the leptonic 
neutralino decays are very sensitive to the CP phases. 
In addition, the production--decay spin correlations lead to 
several CP--even observables such as lepton invariant mass distribution,
and lepton angular distribution, and
one interesting T--odd (CP--odd) triple product of the initial
electron momentum and two final lepton momenta, the size of which might
be large enough to be measured at the high--luminosity future 
electron--positron collider or can play a complementary role in
constraining the CP phases with the EDM constraints.
\end{abstract}

\vskip 0.5cm

\pacs{PACS number(s): 14.80.Ly, 12.60.Jv, 13.85.Qk}


\section{Introduction}
\label{sec:introduction}

The minimal supersymmetric standard model (MSSM) \cite{NHK} is a well--defined
quantum theory of which the Lagrangian form is completely known, including 
the general R--parity preserving, soft supersymmetry (SUSY) breaking terms.
The full MSSM Largarangian has 124 parameters -- 79 real parameters and 45
CP--violating complex phases \cite{DS}. The number of parameters in 
the MSSM is very large compared to 19 in the standard model (SM). 
Therefore, many studies \cite{MS} on possible direct and indirect SUSY 
effects have been made by making several assumptions and investigating 
the variation of a few parameters. Recently, it has, however, been 
shown \cite{BK} that limits on sparticle masses and couplings are very 
sensitive to the assumptions and need to be re-evaluated without making 
any of the simplifying assumptions that have been standard up to now. 

Despite the large number of phases in the model as a whole, only two 
CP-odd rephase-invariant phases, stemming from the chargino
and neutralino mass matrices, take part in the chargino and neutralino
production processes \cite{EHN}. In light of this aspect, phenomenological
analyses with the complex parameter set are not much more difficult 
than those with the real parameter set in the chargino or neutralino 
systems. 
Incidentally, the CP phases are constrained indirectly by the electron 
or neutron electric dipole moment (EDM) and may be small, but the indirect 
constraints \cite{MS} on its actual size depends strongly on the assumptions 
taken in those analyses. As a matter of fact, many recent works \cite{IN} 
have shown that the
constraints could be evaded without suppressing the CP phases of the 
theory. One option \cite{KO} is to make the first two generations of scalar
fermions rather heavy so that one--loop EDM constraints are automatically
evaded. This case can be naturally explained by the so--called effective
SUSY models \cite{DG} where de--couplings of the first and second generation 
sfermions are invoked to solve the SUSY FCNC and CP problems without
spoiling naturalness. Another possibility is to arrange for partial
cancellations among various contributions to the electron and neutron
EDM's. Following the suggestions that the phases do not have to 
be suppressed, many important works on the effects due to the CP phases
have been already reported; the effects are very significant in 
extracting the parameters in the SUSY Lagrangian from experimental 
data \cite{BK}, estimating dark matter densities and scattering cross 
sections \cite{FO} and Higgs boson mass limits \cite{PW}, CP violation in 
the $B$ and $K$ systems \cite{BCKO}, and so on.

If the scale of the SUSY breaking is around 1 TeV as preferred by fine--tuning
arguments in the Higgs sector, many sparticles are expected to be produced
at future colliders such as the Fermilab Tevatron upgrade, the CERN
Large Hadron Collider (LHC), or future $e^+e^-$ colliders proposed by
DESY, KEK and SLAC. Among the sparticles, non--colored supersymmetric
particles such as neutralinos, charginos and sleptons are relatively 
light in most superymmetry theories. With R--parity invariance,
charginos and neutralinos, the mixtures of the gauginos and higgsinos,
are produced pairwise in $e^+e^-$ collisions, either in diagonal or in 
mixed pairs. At LEP2 \cite{LEP2}, and potentially even in 
the first phase of $e^+e^-$  linear colliders (see e.g. Ref.~\cite{NLC}), 
the chargino $\tilde{\chi}_1^\pm$ and the neutralinos 
$\tilde{\chi}^0_{1,2}$
may  be, for some time, the only chargino and neutralino states that can 
be studied experimentally in  detail. Furthermore, as they are expected 
to be lighter than the gluino and in most scenarios lighter than the 
squarks and sleptons, those lighter chargino and neutralino states 
could be first observed in future experiments at $e^+e^-$ colliders.
On the other hand, the heavier chargino and neutralino
states may require the second-phase $e^+e^-$ linear colliders with 
a c.m. energy of about 1.5 TeV.  

In light of the previous generic arguments and aspects concerning the
sparticle spectrum, one of the most promising SUSY processes for 
investigating a wide region of the SUSY parameter space is the 
associated production \cite{MF,BFMM} of neutralinos  
in $e^+e^-$ collisions: $e^+e^-\rightarrow \tilde{\chi}^0_1
\tilde{\chi}^0_2$.
Although in general chargino production \cite{LDK,CMF} is favored by 
larger cross sections, sizable cross sections for the neutralino process
can be expected in certain regions of the parameter space. Moreover,
it might be possible to discover SUSY by neutralino production
if charginos are not accessible.

If neutralinos as new particles are discovered,
its clear identification can be enhanced by utilizing beam polarization
and the complete investigation of the neutralino decays \cite{AM}.
Polarized electron beams have been shown to play a critical role in
disentangling SUSY parameters in chargino--, neutralino-- and 
sfermion--pair productions in $e^+e^-$ collisions. 
However, those works have mainly considered longitudinal electron 
polarization. Recently, an intensive study to obtain high positron 
polarization has been made. In this light, we study the case where 
the polarization of both electron and positron beams can be freely 
manipulated, and we investigate if the highly polarized electron/positron
beams can provide a powerful diagnostic tool for determining SUSY parameters 
in the associated neutralino--pair production.
Angular distributions and angular correlations of the decay products
as well as neutralino decay widths and branching ratios can give valuable 
additional information on their composition from gaugino and higgsino
components. Certainly, one can infer the spin of the new particles
from decay angular distributions with complete spin correlations of the
decaying particle. Moreover, the identification of neutralinos
can be very much solidified by ascertaining the Majorana character
of the neutralinos \cite{BFMM}. This has been demonstrated to be possible 
by means of the energy distributions of the decay leptons if the neutralinos 
are produced in collisions of polarized $e^+e^-$ beams. 
The angular distributions
of the decay products might, however, offer the possibility to prove
the Majorana property although polarized beams are not available.
Furthermore, the angular distributions of the final leptons are suitable
observables for studying CP violation in the MSSM \cite{KOPL}.

After the mixed pair $\tilde{\chi}^0_1$ and $\tilde{\chi}^0_2$
is produced, the second lightest neutralino $\tilde{\chi}^0_2$ 
decays to the LSP and two fermions. Since leptons among fermions
are most cleanly identified at high--performance detectors, one of the 
most promising modes for the associated production and sequential decays
of neutralinos will be  
\begin{center}
\begin{picture}(300,30)(0,0)
\Text(130,25)[]{$e^+e^-\rightarrow\tilde{\chi}^0_1\tilde{\chi}^0_2$}
\Line(155,15)(155,5)
\Text(155,6)[l]{$\rightarrow\tilde{\chi}^0_1+\ell^+\ell^-$}\,.
\end{picture}
\end{center}
So, in the present note we give a comprehensive analysis
through initial beam polarizations and spin correlations between
production and decay to investigate the effects of the CP phases 
and the other SUSY parameters in the associated production and decay 
of the neutralinos in the MSSM with general CP phases but without 
generational mixing. In doing the analysis, it will be
meaningful to include the 
stringent constraints from the electron EDM measurements on the CP phases 
Instead of performing full scans over the phases and real SUSY
parameters, we take two typical scenarios to suppress the electron EDM 
constraints or to allow a large space of the CP phases while
satisfying the electron EDM constraints. The choice of two scenarios
will be made after a global study of the dependence of the electron 
EDM on the relevant parameters in the MSSM. In each scenario, 
the neutralino mass spectrum, three CP--even and one CP--odd observables 
using initial beam polarizations, the neutralino
polarization vector and several distributions observable in the
laboratory frame are presented and discussed.

The organization of the paper is as follows. In Section II 
we describe the supersymmetric flavor--preserving mixing phenomena in a
general parameterization scheme without generation mixing;
selectron left--right mixing, chargino mixing and neutralino mixing, and
identify the relevant CP phases.  
Section III is devoted to the discussion of the constraints by the 
electron EDM on the CP phases and Section IV to the associated production
of neutralinos with polarized beams and the introduction of useful
CP--even and CP--odd observables, which are extracted by controlling the
initial beam polarization. In Section V we describe in detail
the possible decay modes and branching ratios of the second--lightest 
neutralino. In Section VI, we explain how to obtain the fully spin--correlated
distributions of the associated production and decays of the neutralinos
and study the impact of the CP phases on various angular correlations.
Conclusions are given in Section VII.

\section{Supersymmetric Flavor Conserving Mixings}
\label{sec:mixing}

If flavor mixing among sleptons is neglected, the mass matrix of 
selectrons is 
given by
\begin{eqnarray}
 {\cal M}^2_{\tilde{e}}=
 \left(\begin{array}{cc}
       \tilde{m}^2_{\tilde{e}_L}+m^2_e+m^2_Z\cos2\beta\,(s^2_W-1/2) &
       -m_e\left(A^*_e+\mu\tan\beta\right)  \\
       -m_e\left(A_e+\mu^*\tan\beta\right) &
       \tilde{m}^2_{\tilde{e}_R}+m^2_e-m^2_Z\cos2\beta\,s^2_W
       \end{array}
 \right).
\end{eqnarray}
The first term of the diagonal elements is the soft scalar mass term 
evaluated at the weak scale, and the second is the mass squared of the
the corresponding electron (dictated by SUSY), and the last comes from 
the ${\cal D}$ term. The trilinear term $A_e$ causing
left--right mixing is due to the soft--breaking Yukawa--type interaction,
and $\mu$ is the supersymmetric higgsino mass parameter describing
the mixing of two Higgs doublets. And $\tan\beta$ is the ratio $v_2/v_1$
of the vacuum expectation values of the two neutral Higgs fields which
break the electroweak gauge symmetry. The selectron mass eigenstates can be 
obtained by diagonalizing the above mass matrix with a unitary matrix
$U_e$ such that $U_e {\cal M}^2_{\tilde{e}} U^\dagger_e
 = {\rm diag}(m^2_{\tilde{e}_1},m^2_{\tilde{e}_2})$. 
We parameterize $U_e$ so that
\begin{eqnarray}
U_e=\left(\begin{array}{cc}
    \cos\theta_e       &  -\sin\theta_e\,{\rm e}^{-i\phi_e} \\
  \sin\theta_e\,{\rm e}^{i\phi_e}  & \cos\theta_e
          \end{array}\right),
\end{eqnarray}
where $-m_e(A_e+\mu^*\tan\beta)=|m_e(A_e+\mu^*\tan\beta)|{\rm e}^{-i\phi_e}$
and we choose the range of $\theta_e$ and $\phi_e$ so that 
$0\leq\theta_e\leq\pi$ and $-\pi/2\leq\phi_e\leq\pi/2$.

In supersymmetric theories, the spin--1/2 partners of the $W$ bosons and
the charged Higgs bosons, $\tilde{W}^\pm$ and $\tilde{H}^\pm$,
mix to form chargino mass eigenstates $\tilde{\chi}^\pm_{1,2}$. The
chargino mass matrix is given in the $(\tilde{W}^-,\tilde{H}^-)$
basis by
\begin{eqnarray}
{\cal M}_C=\left(\begin{array}{cc}
   M_2                   &    \sqrt{2}m_W\cos\beta     \\
 \sqrt{2}m_W\sin\beta    &     \mu          
                 \end{array}\right),
\end{eqnarray}
which is built up by the fundamental SUSY parameters; the SU(2) gaugino
mass $M_2$ as well as the higgsino mass parameter $\mu$ and the ratio 
$\tan\beta$.
Since the chargino mass matrix ${\cal M}_C$ is not symmetric, two
different unitary matrices acting on the left-- and right--
chiral $(\tilde{W},\tilde{H})$ states are needed to diagonalize the
matrix:
\begin{eqnarray}
U_{L,R}\left(\begin{array}{cc}
             \tilde{W}^-   \\
             \tilde{H}^-
             \end{array}\right)_{L,R}
     = \left(\begin{array}{cc}
             \tilde{\chi}^-_1 \\
             \tilde{\chi}^-_1
             \end{array}\right)_{L,R},
\end{eqnarray}
so that $U_R{\cal M}_CU^\dagger_L={\rm diag}(m_{\tilde{\chi}^\pm_1},
m_{\tilde{\chi}^\pm_2})$ with the ordering $m_{\tilde{\chi}^\pm_1}\leq
m_{\tilde{\chi}^\pm_2}$.

On the other hand, the neutral supersymmetric fermionic partners of 
the $B$ and $W^3$ gauge bosons, $\tilde{B}$ and $\tilde{W}^3$, can mix 
with the neutral supersymmetric fermionic partners of the Higgs bosons, 
$\tilde{H}^0_1$ and $\tilde{H}^0_2$, to form the mass eigenstates. 
Hence the physical states, $\tilde{\chi}^0_i$, called neutralinos, are 
found by diagonalizing the $4\times 4$ mass matrix
\begin{eqnarray}
{\cal M}_N=\left(\begin{array}{cccc}
          M_1                   &              0                     
 &   -m_Zc_\beta s_W  &   m_Zs_\beta s_W \\
           0                    &             M_2         
 &    m_Zc_\beta c_W  &  -m_Zs_\beta c_W \\
     -m_Zc_\beta s_W  &   m_Zc_\beta s_W
 &         0                    &            -\mu            \\ 
      m_Zs_\beta s_W  &  -m_Zs_\beta c_W
 &        -\mu                  &              0
                \end{array}\right),
\label{eq:neutralino mass matrix}
\end{eqnarray}
where $s_\beta=\sin\beta$, $c_\beta=\cos\beta$, and $s_W, c_W$ are the sine 
and cosine of the electroweak mixing angle $\theta_W$, respectively.
The neutralino mass matrix ${\cal M}_N$ is a complex, symmetric
matrix so that it can be diagonalized by just one unitary matrix $N$ such that
$ N^*{\cal M}_N N^\dagger = {\rm diag}(m_{\tilde{\chi}^0_1},
m_{\tilde{\chi}^0_2}, m_{\tilde{\chi}^0_3},m_{\tilde{\chi}^0_4})$ with the 
ordering
$m_{\tilde{\chi}^0_1}\leq m_{\tilde{\chi}^0_2}\leq m_{\tilde{\chi}^0_3}
\leq m_{\tilde{\chi}^0_4}$.

In CP--noninvariant theories, the gaugino mass $M_2, M_1$ and the 
higgsino mass parameter $\mu$ as well as the trilinear parameter $A_e$
can be complex. However, by reparametrization of the
fields, $M_2$ can be assumed real and positive without loss of
generality since all other parameter choices are related to our choice
by an appropriate $R$ transformation. 
Taking into account our parameterization choice, the final
set of phases considered in the discussion of the electron EDM and the
neutralino production and decays includes two phases appearing in the
chargino--neutralino sector $\Phi_1, \Phi_\mu$ and one phase $\Phi_{A_e}$
corresponding to the trilinear soft breaking parameter relevant in the
electric dipole moment calculation as will be discussed in the following 
section:
\begin{eqnarray}
\mu=|\mu|\,{\rm e}^{i\Phi_\mu},~~~ M_1=|M_1|\,{\rm e}^{i\Phi_1},~~~
A_e=|A_e|\,{\rm e}^{i\Phi_{A_e}}.
\end{eqnarray}
We note that even though the off-diagonal elements of the
selectron mass matrix are proportional to the small electron Yukawa coupling,
the CP phases, in particular, $\Phi_{A_e}$, play a crucial role in determining
the size of the electron EDM because every SUSY contribution to the EDM 
requires a chirality flip leading to dipole moments' proportionality to 
the electron mass $m_e$.

\section{Electric dipole moment of the electron}
\label{sec:electron EDM}

The electric dipole interaction of a spin--1/2 electron $e$ with an 
electromagnetic field is described by an effective Lagrangian
\begin{eqnarray}
{\cal L}_{EDM}=-\frac{i}{2}d_e\,\bar{e}\sigma^{\mu\nu}\gamma_5 e\, F_{\mu\nu}.
\end{eqnarray}
In theories with CP--violating interactions, the electric dipole moment 
$d_e$ receives contributions from one loop diagrams. In the MSSM, two diagrams
contribute to the electron EDM in the mass eigenstate basis of all
particles. They are shown in Fig.~1 (summation over all charginos
and neutralinos in the loops is understood).

\vspace*{1cm} 
\begin{center}
\begin{picture}(330,110)(0,0)

\Text(10,90)[]{$e$}
\ArrowLine(20,90)(50,90)
\Text(75,102)[]{$\tilde{\nu}_e$}
\Line(50,89.5)(100,89.5)
\Line(50,90.5)(100,90.5)
\ArrowLine(100,90)(130,90)
\Text(140,90)[]{$e$}
\Line(50,90)(75,50)
\Photon(50,90)(75,50){4}{7}
\Text(50,65)[]{$\tilde{\chi}^-_j$}
\Line(75,50)(100,90)
\Photon(75,50)(100,90){4}{7}
\Text(100,65)[]{$\tilde{\chi}^-_j$}
\Photon(75,50)(75,20){4}{7}
\Text(75,12)[]{$\gamma$}
\Text(75,0)[]{${\rm (a)}$}

\Text(170,90)[]{$e$}
\ArrowLine(180,90)(210,90)
\ArrowLine(260,90)(290,90)
\Text(235,105)[]{$\tilde{\chi}^0_j$}
\Line(210,90)(260,90)
\Photon(210,90)(260,90){4}{7}
\Text(300,90)[]{$e$}
\Line(210,89)(235,49)
\Line(210,91)(235,51)
\Text(210,65)[]{$\tilde{e}_a$}
\Line(235,49)(260,89)
\Line(235,51)(260,91)
\Text(260,65)[]{$\tilde{e}_a$}
\Photon(235,50)(235,20){4}{7}
\Text(235,12)[]{$\gamma$}
\Text(235,0)[]{${\rm (b)}$}

\end{picture}\\
\end{center}

\noindent
Figure~1: {\it Diagrams contributing to the electron EDM; 
               (a) chargino--exchange contributions and 
               (b) neutralino--exchange contributions.}

\vskip 0.5cm

The Lagrangian describing the $\tilde{\chi}^\pm$--$e$-$
\tilde{\nu}_e$ interactions without flavor mixing is
\begin{eqnarray}
{\cal L}_{\tilde{\chi}^\pm e\tilde{\nu}} = \frac{e}{s_W}
 \bar{e}\left(Y_e U^*_{Lj2}P_L - U^*_{Rj1}P_R\right)
\tilde{\chi}^-_j\tilde{\nu}_e
 +{\rm h.c.},
\end{eqnarray}
where $P_{L,R}=(1\mp\gamma_5)/2$,
and the interaction Lagrangian describing the most general 
$\tilde{\chi}^0$-$e$-$\tilde{e}$ interactions are given in terms of 
mass eigenstates by
\begin{eqnarray}
{\cal L}_{\tilde{\chi}^0e\tilde{e}} = -\frac{e}{\sqrt{2}s_W}\bar{e}
   \left[B^{aL}_j P_L+B^{aR}_j P_R\right]\tilde{\chi}^0_j\,\tilde{e}_a
   +{\rm h.c.},
\end{eqnarray}
where $a=1,2$ and the couplings $B^{aL}_j$ and $B^{aR}_j$ are given by
\begin{eqnarray}
&& B^{1L}_j = \sqrt{2}Y_e N^*_{j3}\cos\theta_e
             +2N^*_{j1}\tan\theta_W{\rm e}^{i\phi_e}\sin\theta_e\,,
             \nonumber\\
&& B^{1R}_j = -(N_{j2}+N_{j1}\tan\theta_W)\cos\theta_e
             +\sqrt{2}Y_e N_{j3}{\rm e}^{i\phi_e}\sin\theta_e\,,\nonumber\\
&& B^{2L}_j = -\sqrt{2}Y_e N^*_{j3}{\rm e}^{-i\phi_e}\sin\theta_e
             +2N^*_{j1}\tan\theta_W\cos\theta_e\,,\nonumber\\
&& B^{2R}_j = (N_{j2}+N_{j1}\tan\theta_W){\rm e}^{-i\phi_e}
             +\sqrt{2}Y_e N_{j3}\cos\theta_e\,,
\end{eqnarray}
with the electron Yukawa coupling $Y_e=m_e/(\sqrt{2}m_W \,c_\beta)\approx
6.4\times 10^{-6}/c_\beta$.

It is clear that the matrix elements of two unitary matrices $U_L$ and $U_R$ 
diagonalizing the chargino mass matrix ${\cal M}_C$ are functions
of the phase $\Phi_\mu$ but not of the phase $\Phi_1$. 
Using the chargino--electron--sneutrino interaction,
we find that the chargino contribution to the EDM for the electron
through the diagram shown in Fig.~1(a) is 
\begin{eqnarray}
\frac{1}{e}d^{\tilde{\chi}^\pm}_e = \frac{\alpha}{4\pi s^2_W}
   Y_e\sum_{j=1}^2\left\{\frac{m_{\tilde{\chi}^\pm_j}}{m^2_{\tilde{\nu}}}
   {\cal I}\left[U^*_{Lj2}U_{Rj1}\right]
   A\left(\frac{m^2_{\tilde{\chi}^\pm_j}}{m^2_{\tilde{\nu}}}\right)
   \right\},
\end{eqnarray}
where $A(r)=2(1-r)^{-2}[3-r+2\,{\rm ln}r(1-r)^{-1}]$. On the other hand,
the neutralino diagonalization matrix $N$ is a function of both $\Phi_1$ 
and $\Phi_\mu$. Using the neutralino--electron--selectron interaction,
we find that the neutralino contribution to the EDM of the electron
through the diagram shown in Fig.~2(b) is 
\begin{eqnarray} 
 \frac{1}{e}d^{\tilde{\chi}^0}_e =-\frac{\alpha}{8\pi s^2_W}
   \sum_{a=1}^2\sum_{j=1}^4
   \left\{\frac{m_{\tilde{\chi}^0_j}}{m^2_{\tilde{e}_a}}
   {\cal I}\left[B^{aL}_j B^{aR*}_j\right]
   B\left(\frac{m^2_{\tilde{\chi}^\pm_j}}{m^2_{\tilde{e}_a}}\right)
   \right\},
\end{eqnarray}
where $B(r)=2(1-r)^{-2}[1+r+2\,{\rm ln}r(1-r)^{-1}]$. Our anlaytical 
expression for the electron EDM is consistent with that of Pokorski, 
Rosiek and Savoy of Ref.~\cite{IN} although there is a small difference
in the neutralino contribution between our result and that of Brhlik, 
Good and Kane \cite{IN}, of which the expression (A10) must have a 
negative sign 
in the last term instead of a positive sign. These results are completely 
general except for flavor mixing and lead to the MSSM contribution 
to the electron EDM as the sum of two contributions:
\begin{eqnarray}
d_e=d^{\tilde{\chi}^\pm}_e+d^{\tilde{\chi}^0}_e\,.
\end{eqnarray}
Of course, the Kobayashi--Maskawa CP phase in the SM can in principle 
contribute to the electron EDM, but it turns out to be effective
only at three--loop level so that the contribution is too small to be 
measured.

One of the important features of the SUSY contributions to the electron EDM
is the fact that the EDM requires different chirality of the initial
and final electrons. In the supersymmetric diagrams this chirality flip can 
happen in two ways -- either the exchanged selectrons change chirality via
$L$-$R$ mixing terms in the selectron mass matrix and couple to the
gaugino component of the intermediate spin--1/2 particle, or the left-- and
right--handed selectrons/sneutrinos preserve their chirality and couple to the
higgsino components of the charginos or neutralinos, respectively.
As a result, all contributions are directly proportional to the mass
of the external electron since both the $L$-$R$ mixing selectron mass 
term and the Higgsino--electron-selectron (or sneutrino) coupling are 
proportional to the relevant electron Yukawa coupling $Y_e$. 
Another consequence of the chirality flip is the explicit proportionality
of the contributions to the mass of the intermediate spin--1/2 particle.

Generally, the SUSY contributions to the electron EDM are determined by 7 real 
parameters $\{\tan\beta, |M_1|,M_2,|\mu|,m_{\tilde{e}_L},m_{\tilde{e}_R},
|A_e|\}$ and 3 CP phases $\{\Phi_1,\Phi_\mu,\Phi_{A_e}\}$. So, in order to 
understand the general features of the SUSY contribution effectively,
it will be necessary to make some appropriate specifications without 
spoiling their qualitative aspects. We take a universal soft--breaking 
selectron mass $m_{\tilde{e}}$ for the left-- and right--handed selectrons,
because at any rate two states are not degenerate due to extra 
contributions to their masses. For six real SUSY parameters, we consider 
two typical scenarios where three CP phases are left as free parameters.  
In both scenarios, the gaugino mass unification condition 
is assumed only for the modulus of 
the gaugino mass parameters; $|M_1|=\frac{5}{3}\tan^2\theta_W\,M_2
\approx 0.5\,M_2$ and the size of the trilinear parameter $|A_e|$ is set 
to 1 TeV through the paper. 
It is necessary to be careful in choosing the value of 
$\tan\beta$, according to which various physical quantities will be very
different. The recent calculation by Chang, Keung and Pilaftsis \cite{CKP} 
for the Barr--Zee--type two--loop contributions to the EDM's, 
the sbottom contributions are very much enhanced for a large $\tan\beta$
\cite{BCDP} so that the contributions cannot be simply neglected. 
Therefore, for 
a large value of $\tan\beta$ we are forced to introduce more CP phases
related with sparticles of the third generation in our analysis. 
Furthermore, as will be discussed in more detail in the section for the 
neutralino decays, for a large $\tan\beta$ we need 
to include stau left--right mixing as well as Higgs--exchange diagrams in 
evaluating different branching fractions of neutralino decays. 
On the contrary, the value of $\tan\beta\leq 2.5$ has been
already ruled out by null results in the Higgs search experiments at LEP II
\cite{HIGGS}.
Postponing the detailed analyses related with the $\tan\beta$ dependence of
the EDM's, the associated production of neutralinos and the  branching ratios 
of neutralino decays to our next work, we simply take $\tan\beta=3$ in the 
present analysis and treat the stau contributions on the same footing as
the other slepton contributions. 

With these several specifications on the SUSY parameters, the electron EDM is 
determined by three real parameters $\{M_2,|\mu|,m_{\tilde{e}}\}$ and three 
remaining 
phases $\{\Phi_1,\Phi_\mu,\Phi_{A_e}\}$. The first scenario ${\cal S}1$, 
which is based on the so-called effective SUSY model, decouples selectrons 
by rendering them extremely heavy without violating the naturalness
arguments, but taking $M_2$ and $|\mu|$ relatively small:
\begin{eqnarray}
{\cal S}1:\ \ M_2=100\,{\rm GeV},\ \ |\mu|=200\,{\rm GeV},\ \ 
m_{\tilde{e}}=10\,{\rm TeV},
\end{eqnarray}
where 10 TeV for $m_{\tilde{e}}$ is taken because it is
large enough to suppress the selectron
contributions (almost) completely. As a result, the present electron EDM 
measurements \cite{eEDM} of $|d_e|\leq 4.3\times 10^{-27}\,e\cdot{\rm cm}$
do not put any constraints on the CP phases.
The second scenario ${\cal S}2$ takes a small universal soft--breaking
selectron mass, but a large value of $|\mu|$:
\begin{eqnarray}
{\cal S}2:\ \ M_2=100\,{\rm GeV},\ \ |\mu|=700\,{\rm GeV},\ \
m_{\tilde{e}}=200\,{\rm GeV}.
\end{eqnarray}
In this scenario, we expect that some cancellations among the
CP phases are needed to suppress the electron EDM. Of course, the
degree of the cancellations depends on the values of the real SUSY
parameters, especially the higgsino mass parameter $|\mu|$. 
The reason why we take a large $|\mu|$ of 700 GeV is to allow a relatively
large region for the CP phases $\Phi_\mu$ and $\Phi_1$ while scanning the
phase $\Phi_{A_e}$ from 0 to $2\pi$. The fact that the allowed region of
two phases increases with $|\mu|$ has been pointed out in the work
by Brhlik, Good and Kane \cite{IN}. 

We display in Figure~2 (a) the allowed range of $\Phi_\mu$ versus 
$|\mu|$ at 95\% confidence level in the scenario ${\cal S}2$
for other phases sampled randomly within their allowed ranges.
The overall trend clearly shows that for larger values of $|\mu|$ it is much 
easier to satisfy the electron EDM limits and any value of $|\mu|$ larger
than 650 GeV allows the full range of $\Phi_\mu$. 
Figure~2(b) shows the allowed region at 95\% confidence level for the
phases $\Phi_\mu$ and $\Phi_1$ in the scenario ${\cal S}2$ with $|\mu|=
700$ GeV. Note that near the region for $\Phi_1=\pi$ the phase 
$\Phi_\mu$ can take any value. In our next analysis on the correlated 
associated production and decay of neutralinos, which are mainly
related with the CP phases $\Phi_\mu$ and $\Phi_1$ but not with the phase 
$\Phi_{A_e}$, Figure~2(b) will serve as the basic platform for all the 
contour plots for production cross sections, total cross 
sections of the correlated process, the branching ratios, several CP--even 
and CP--odd observables and an interesting CP--odd (T--odd) triple momentum 
product of the initial electron momentum and two lepton momenta from the
leptonic decays of the neutralinos.

\section{Associated Production of Neutralinos}
\label{sec:neutralino pair productions}

\subsection{Production helicity amplitudes}

Although we are mainly interested in one production process 
$e^+e^-\rightarrow\tilde{\chi}^0_2\tilde{\chi}^0_1$, we discuss
in this section the associated production of every combination of
neutralino--pair $e^+e^-\rightarrow\tilde{\chi}^0_i\tilde{\chi}^0_j$ 
[$i,j=1$--4] on a general footing. Note that the chirality mixing of 
scalar electrons are determined by the very small electron Yukawa coupling 
proportional to the electron mass much smaller than the collider 
c.m. energy (500 GeV) under consideration by a factor of about 10$^{6}$. 
Therefore, the selectron left--right chirality mixing can be safely 
neglected in the associated production of neutralinos so that the trilinear
term $A_e$ does not play any role in the high energy process
unlike the case for the electron EDM. In this approximation, the production 
process $e^+e^-\rightarrow
\tilde{\chi}^0_i\tilde{\chi}^0_j$ is generated by the five mechanisms 
shown in Fig.~3: $s$-channel $Z$ exchange, $t$-channel $\tilde{e}_{L,R}$ 
exchanges, and $u$-channel $\tilde{e}_{L,R}$ exchanges. 
The transition matrix element, after an appropriate Fierz
transformation of the $\tilde{e}_{L,R}$ exchange amplitudes
\begin{eqnarray}
T\left(e^+e^-\rightarrow\tilde{\chi}^0_i\tilde{\chi}^0_j\right)
 = \frac{e^2}{s}Q^{ij}_{\alpha\beta}
   \left[\bar{v}(e^+)  \gamma_\mu P_\alpha  u(e^-)\right]
   \left[\bar{u}(\tilde{\chi}^0_i) \gamma^\mu P_\beta 
               v(\tilde{\chi}^0_j)\right]\,,
\label{eq:neutralino production amplitude}
\end{eqnarray}
can be expressed in terms of four generalized bilinear charges, classified 
according to the chiralities $\alpha,\beta=L,R$ of the associated electron
and neutralino currents
\begin{eqnarray}
Q^{ij}_{LL}&=&+\frac{D_Z}{s_W^2c_W^2}(s_W^2 -\frac{1}{2}){\cal Z}_{ij}
              -D_{uL}g_{Lij},\nonumber\\ 
Q^{ij}_{LR}&=&-\frac{D_Z}{s_W^2c_W^2}(s_W^2 -\frac{1}{2}){\cal Z}^*_{ij}
              +D_{tL}g^*_{Lij},\nonumber\\
Q^{ij}_{RL}&=&+\frac{D_Z}{c_W^2}{\cal Z}_{ij}
              +D_{tR}g_{Rij},\nonumber\\ 
Q^{ij}_{RR}&=&-\frac{D_Z}{c_W^2}{\cal Z}^*_{ij}
              -D_{uR}g^*_{Rij},
\end{eqnarray}
with $s$--, $t$--, and $u$--channel propagators and the couplings 
$Z_{ij}$, $g_{Lij}$ and $g_{Rij}$:
\begin{eqnarray}
&& D_Z=\frac{s}{s-m^2_Z+im_Z\Gamma_Z},\nonumber\\
&& D_{tL,R}=\frac{s}{t-m^2_{\tilde{e}_{L,R}}},\nonumber\\
&& D_{uL,R}=\frac{s}{u-m^2_{\tilde{e}_{L,R}}},
\end{eqnarray}
with $s=(p_e+p_{\bar{e}})^2$, $t=(p_e-p_{\tilde{\chi}^0_i})^2$ and
$u=(p_e-p_{\tilde{\chi}^0_j})^2$. And, the combinations ${\cal Z}_{ij}$, 
$g_{Lij}$ and $g_{Rij}$ of the neutralino diagonalization matrix elements
$N_{ij}$
\begin{eqnarray}
&& {\cal Z}_{ij}=\frac{1}{2}
                 \left[N_{i3}N^*_{j3}-N_{i4}N^*_{j4}\right]\,,\nonumber\\
&& g_{Lij}=\frac{1}{4 s_W^2c_W^2}(N_{i2}c_W+N_{i1}s_W)(N^*_{j2}c_W+N^*_{j1}s_W)
                 \,,\nonumber\\
&& g_{Rij}=\frac{1}{c_W^2}N_{i1}N^*_{j1}\,,
\end{eqnarray}
satisfy the hermiticity relations reflecting the CP relations 
\begin{eqnarray}
{\cal Z}_{ij}={\cal Z}^*_{ji},\qquad
g_{Lij}=g^*_{Lji},\qquad
g_{Rij}=g^*_{Rji},
\end{eqnarray}
so that, if the $Z$--boson width $\Gamma_Z$ is neglected in the 
$Z$--boson propagator $D_Z$, the bilinear charges $Q^{ij}_{\alpha\beta}$ 
also satisfy the same relations $Q^{ij}_{\alpha\beta}=Q^{ji*}_{\alpha\beta}$
with $t$ and $u$ interchanged in the propagators. The relation
is very useful in classifying CP--even and CP--odd observables
in the following.

\vspace*{1cm} 
\begin{center}
\begin{picture}(330,100)(0,0)

\Text(15,85)[]{$e^-$}
\ArrowLine(10,75)(35,50)
\ArrowLine(35,50)(10,25)
\Text(15,15)[]{$e^+$}
\Photon(35,50)(75,50){4}{8}
\Text(55,37)[]{$Z$}
\Line(75,50)(100,75)
\Photon(75,50)(100,75){3}{7}
\Text(97,87)[]{$\tilde{\chi}^0_i$}
\Line(100,25)(75,50)
\Photon(100,25)(75,50){3}{7}
\Text(97,13)[]{$\tilde{\chi}^0_j$}

\Text(125,85)[]{$e^-$}
\ArrowLine(120,75)(165,75)
\Text(125,15)[]{$e^+$}
\ArrowLine(165,25)(120,25)
\Line(164,75)(164,25)
\Line(166,75)(166,25)
\Text(153,50)[]{$\tilde{e}_{L,R}$}
\Line(165,75)(210,75)
\Photon(165,75)(210,75){3}{7}
\Text(207,87)[]{$\tilde{\chi}^0_i$}
\Line(210,25)(165,25)
\Photon(210,25)(165,25){3}{7}
\Text(207,13)[]{$\tilde{\chi}^0_j$}

\Text(235,85)[]{$e^-$}
\ArrowLine(230,75)(275,75)
\Text(235,15)[]{$e^+$}
\ArrowLine(275,25)(230,25)
\Line(274,75)(274,25)
\Line(276,75)(276,25)
\Text(260,50)[]{$\tilde{e}_{L,R}$}
\Line(275,75)(320,25)
\Photon(275,75)(320,25){3}{7}
\Text(317,87)[]{$\tilde{\chi}^0_i$}
\Line(320,75)(275,25)
\Photon(320,75)(275,25){3}{7}
\Text(317,13)[]{$\tilde{\chi}^0_j$}
\end{picture}\\
\end{center}

\noindent
Figure~3: {\it Five mechanisms contributing to the production of 
               neutralino pairs in $e^+e^-$ annihilation, 
               $e^+e^-\rightarrow \tilde{\chi}^0_i \tilde{\chi}^0_j$.}
\vskip 0.5cm

All physical observables (which can be constructed and measured through 
the production process) are expressed in a simple form by 16 so--called 
quartic charges \cite{SZ} which contain important dynamical properties of 
the process and are expressed in terms of the bilinear charges
$Q^{ij}_{\alpha\beta}$. These quartic charges are classified according to 
their transformation properties under parity as follows:\\
\vskip 0.2cm
\noindent
{\bf (a)} \underline{Eight P--even terms}:
\begin{eqnarray}
&& Q^{ij}_1=\frac{1}{4}
            \left[|Q^{ij}_{RR}|^2+|Q^{ij}_{LL}|^2
                 +|Q^{ij}_{RL}|^2+|Q^{ij}_{LR}|^2\right]\,,\nonumber\\
&& Q^{ij}_2=\frac{1}{2}{\cal R}
            \left[Q^{ij}_{RR}Q^{ij*}_{RL}
                 +Q^{ij}_{LL}Q^{ij*}_{LR}\right]\,,\nonumber\\
&& Q^{ij}_3=\frac{1}{4}
            \left[|Q^{ij}_{LL}|^2+|Q^{ij}_{RR}|^2
                 -|Q^{ij}_{RL}|^2-|Q^{ij}_{LR}|^2\right]\,,\nonumber\\
&& Q^{ij}_4=\frac{1}{2}{\cal I}
            \left[Q^{ij}_{RR}Q^{ij*}_{RL}
                 +Q^{ij}_{LL}Q^{ij*}_{LR}\right]\,,\nonumber\\
&& Q^{ij}_5=\frac{1}{2}{\cal R}
            \left[Q^{ij}_{RR}Q^{ij*}_{LR}
                 +Q^{ij}_{LL}Q^{ij*}_{RL}\right]\,,\nonumber\\
&& Q^{ij}_6=\frac{1}{2}{\cal I}
            \left[Q^{ij}_{RR}Q^{ij*}_{LR}
                 +Q^{ij}_{LL}Q^{ij*}_{RL}\right]\,,\nonumber\\
&& Q^{ij}_7={\cal R}\left[Q^{ij}_{RR}Q^{ij*}_{LL}\right]\,,\nonumber\\
&& Q^{ij}_8={\cal R}\left[Q^{ij}_{RL}Q^{ij*}_{LR}\right]\,,
\end{eqnarray}\\
{\bf (b)} \underline{Eight P--odd terms}: 
\begin{eqnarray}
&& Q^{'ij}_1=\frac{1}{4}
             \left[|Q^{ij}_{RR}|^2+|Q^{ij}_{RL}|^2
                  -|Q^{ij}_{LR}|^2-|Q^{ij}_{LL}|^2\right]\,,\nonumber\\
&& Q^{'ij}_2=\frac{1}{2}{\cal R}
             \left[Q^{ij}_{RR}Q^{ij*}_{RL}
                  -Q^{ij}_{LL}Q^{ij*}_{LR}\right]\,,\nonumber\\
&& Q^{'ij}_3=\frac{1}{4}
             \left[|Q^{ij}_{RR}|^2+|Q^{ij}_{LR}|^2
                  -|Q^{ij}_{RL}|^2-|Q^{ij}_{LL}|^2\right]\,,\nonumber\\
&& Q^{'ij}_4=\frac{1}{2}{\cal I}
             \left[Q^{ij}_{RR}Q^{ij*}_{RL}
                  -Q^{ij}_{LL}Q^{ij*}_{LR}\right]\,,\nonumber\\
&& Q^{'ij}_5=\frac{1}{2}{\cal R}
             \left[Q^{ij}_{RR}Q^{ij*}_{LR}
                  -Q^{ij}_{LL}Q^{ij*}_{RL}\right]\,,\nonumber\\
&& Q^{'ij}_6=\frac{1}{2}{\cal I}
             \left[Q^{ij}_{RR}Q^{ij*}_{LR}
                  -Q^{ij}_{LL}Q^{ij*}_{RL}\right]\,,\nonumber\\
&& Q^{'ij}_7={\cal I}\left[Q^{ij}_{RR}Q^{ij*}_{LL}\right]\,,\nonumber\\
&& Q^{'ij}_8={\cal I}\left[Q^{ij}_{RL}Q^{ij*}_{LR}\right]\,.
\end{eqnarray}
We note that these 16 quartic charges comprise the most complete set for
any fermion--pair production process in $e^+e^-$ collisions when the electron 
mass is neglected. On the other hand, the quartic charges defined by an 
imaginary
part of the bilinear--charge correlations might be nonvanishing only when 
there are complex CP--violating couplings or/and CP--preserving phases 
like recattering phases or finite widths of the intermediate particles. 
So, if there are no CP--preserving phases, non--vanishing values 
of these quartic charges signal CP violation in the given process.

Defining the $\tilde{\chi}^0_i$ production angle with respect to the
electron flight direction by $\Theta$, the helicity amplitudes can be 
determined from Eq.~(\ref{eq:neutralino production amplitude}). 
Electron and positron helicities are opposite to each other in 
all exchange amplitudes, but the $\tilde{\chi}^0_i$ and $\tilde{\chi}^0_j$ 
helicities are less correlated due to the non--zero masses of the 
particles; amplitudes with equal neutralino helicities 
must vanish only $\propto m_{\tilde{\chi}^0_{i,j}}/\sqrt{s}$ for asymptotic 
energies. Denoting the electron helicity by the first index, 
the $\tilde{\chi}^0_i$ and $\tilde{\chi}^0_j$ helicities by the remaining 
two indices, the helicity amplitudes $T(\sigma;\lambda_i,
\lambda_j)=2\pi\alpha\langle\sigma;\lambda_i\lambda_j\rangle$ are 
given by
\begin{eqnarray}
&& \langle +;++\rangle 
   =-\left[Q^{ij}_{RR}\sqrt{1-\eta^2_+}+Q^{ij}_{RL}\sqrt{1-\eta^2_-}\right]
     \sin\Theta\,,\nonumber\\
&& \langle +;+-\rangle 
   =-\left[Q^{ij}_{RR}\sqrt{(1+\eta_+)(1+\eta_-)}+
           Q^{ij}_{RL}\sqrt{(1-\eta_+)(1-\eta_-)}\right](1+\cos\Theta)
               \,,\nonumber\\
&& \langle +;-+\rangle 
   =+\left[Q^{ij}_{RR}\sqrt{(1-\eta_+)(1-\eta_-)}+
           Q^{ij}_{RL}\sqrt{(1+\eta_+)(1+\eta_-)}\right](1-\cos\Theta)
              \,, \nonumber\\
&& \langle +;--\rangle 
   =+\left[Q^{ij}_{RR}\sqrt{1-\eta^2_-}+Q^{ij}_{RL}\sqrt{1-\eta^2_+}\right]
     \sin\Theta\,, 
\end{eqnarray}
for the right--handed electron beam, and
\begin{eqnarray}
&& \langle -;++\rangle 
   =-\left[Q^{ij}_{LR}\sqrt{1-\eta^2_+}+Q^{ij}_{LL}\sqrt{1-\eta^2_-}\right]
     \sin\Theta \,,\nonumber\\
&& \langle -;+-\rangle 
   =+\left[Q^{ij}_{LR}\sqrt{(1+\eta_+)(1+\eta_-)}+
           Q^{ij}_{LL}\sqrt{(1-\eta_+)(1-\eta_-)}\right](1-\cos\Theta)
               \,,\nonumber\\
&& \langle -;-+\rangle 
   =-\left[Q^{ij}_{LR}\sqrt{(1-\eta_+)(1-\eta_-)}+
           Q^{ij}_{LL}\sqrt{(1+\eta_+)(1+\eta_-)}\right](1+\cos\Theta)
              \,, \nonumber\\
&& \langle -;--\rangle 
   =+\left[Q^{ij}_{LR}\sqrt{1-\eta^2_-}+Q^{ij}_{LL}\sqrt{1-\eta^2_+}\right]
     \sin\Theta\,, 
\label{eq:helicity amplitude}
\end{eqnarray}
for the left--handed electron beam,
where $\eta_\pm=\lambda^{1/2}(1,\mu^2_i,\mu^2_j)\pm(\mu^2_i-\mu^2_j)$
with $\mu^2_j=m^2_{\tilde{\chi}^0_i}/s$ and 
$\lambda(x,y,z)=x^2+y^2+z^2-2xy-2yz-2zx$. The explicit form of the production
helicity amplitudes have been obtained by the so--called 2--component spinor
technique of Ref.~\cite{HZ}. If the arguments are not specified, 
the notation $\lambda$ stands for $\lambda(1,\mu^2_i,\mu^2_j)$ in the 
following. 

\subsection{Neutralino mass spectrum and production cross section}
\label{sec:production cross section}

\subsubsection{Neutralino masses}

Before investigating various dynamical distributions in the neutralino
processes, it will be worthwhile to see the dependence of the neutralino 
masses and of the gaugino composition of the two light neutralino states 
on the CP phases in two scenarios 
${\cal S}1$ and ${\cal S}2$.  Figure~4 shows the mass spectrum of 
the neutralinos $\tilde{\chi}^0_{1,2}$ on the plane of the CP phases
$\Phi_\mu$ and $\Phi_1$ in the two scenarios; 
(a) ${\cal S}1$ (upper figures) and (b) ${\cal S}2$ (lower figures).
Except for the region around $\Phi_\mu=0,2\pi$ 
in ${\cal S}1$, the second--lightest neutralino mass $m_{\tilde{\chi}^0_2}$ 
is (almost) independent of $\Phi_1$ in both ${\cal S}1$ and ${\cal S}2$, while 
the lightest neutralino mass $m_{\tilde{\chi}^0_1}$ exhibits a very strongly
correlated dependence on the CP phases. Note that $m_{\tilde{\chi}^0_1}$
becomes maximal at non--trivial values of $\Phi_\mu$ and $\Phi_1$
in ${\cal S}1$ with $|\mu|=200$ GeV. This feature leads immediately to the
conclusion that $m_{\tilde{\chi}^0_1}$ is strongly affected by a small value 
of $|\mu|$, while $m_{\tilde{\chi}^0_2}$ is essentially determined by the 
SU(2) gaugino mass $M_2$. Combined with the electron EDM constraints shown as
the shadowed region in Fig.~1, the mass $m_{\tilde{\chi}^0_1}$
becomes smaller as the CP phases $\Phi_1$ and $\Phi_\mu$ approach the
off--diagonal line on the plane, which implies that the
mass is a function of the sum $\Phi_\mu+\Phi_1$ of two CP phase to a very 
good approximation.

Since the phase $\Phi_1$ is related with the gaugino part while the phase
$\Phi_\mu$ with the higgsino part, the size of their contributions will be
strongly dependent on the size of the gaugino (or higgsino) compositions 
of the neutralino states. So, we present in Figure~5 the gaugino compositions 
${\cal X}_1$ and ${\cal X}_2$ of the lightest and second--lightest 
neutralinos $\tilde{\chi}^0_1$ and $\tilde{\chi}^0_2$ defined by 
\begin{eqnarray}
{\cal X}_1 = |N_{11}|^2+|N_{12}|^2,\nonumber\\
{\cal X}_2 = |N_{21}|^2+|N_{22}|^2,
\end{eqnarray}
with respect to the phase $\Phi_\mu$ while the phase $\Phi_1$ is 
scanned over its full allowed range in the scenarios (a) ${\cal S}1$
and (b) ${\cal S}2$. As expected, $\tilde{\chi}^0_1$ has larger gaugino
composition than $\tilde{\chi}^0_2$. Certainly, the gaugino composition
in the scenario ${\cal S}2$ is almost 100\% due to the large
value of $|\mu|=700$ GeV compared to the gaugino masses $|M_1|$ and
$M_2$. For $\Phi_\mu$ around $\pi$, the gaugino compositions are almost 
insensitive to the phase $\Phi_1$. 
On the contrary, in the region of 
$\Phi_\mu=0,2\pi$, the gaugino compositions are very sensitive to the phase 
$\Phi_1$. This feature is partially responsible for the fact that in the 
scenario ${\cal S}1$, the neutralino masses are strongly dependent on the 
phase $\Phi_1$ around $\Phi_\mu=0,2\pi$.

\subsubsection{Production cross section}

One of the important distributions in the production process 
$e^+e^-\rightarrow\tilde{\chi}^0_i\tilde{\chi}^0_j$ is the differential 
cross section averaged 
over the initial beam polarizations. This unpolarized differential production 
cross section is given by taking the average/sum over the initial/final 
helicities:
\begin{eqnarray}
\frac{{\rm d}\sigma}{{\rm d}\cos\Theta}
      (e^+e^-\rightarrow\tilde{\chi}^0_i\tilde{\chi}^0_j)
 =\frac{\pi\alpha^2}{32 s} \lambda^{1/2} \, 
  \sum |\langle\sigma;\lambda_i\lambda_j\rangle|^2\,.
\end{eqnarray}
Carrying out the sum, one finds the following expression for the 
differential cross section in terms of the scattering angle
$\Theta$ and the quartic charges:
\begin{eqnarray}
\frac{{\rm d}\sigma}{{\rm d}\cos\Theta}
      (e^+e^-\rightarrow\tilde{\chi}^0_i\tilde{\chi}^0_j)
&=&\frac{\pi\alpha^2}{8 s} \lambda^{1/2} 
  \bigg\{[4-(\eta_+-\eta_-)^2+(\eta_++\eta_-)^2\cos^2\Theta]Q^{ij}_1\,,
       \nonumber\\
&&+4\sqrt{(1-\eta^2_+)(1-\eta^2_-)}Q^{ij}_2
             +4(\eta_++\eta_-)\cos\Theta Q^{ij}_3\bigg\}\,.
\label{eq:cross section}
\end{eqnarray}
Figure~6 shows the dependence of the production cross section on the
scattering angle $\Theta$ and on the CP phases $\{\Phi_\mu,\Phi_1\}$
in the scenario (a) ${\cal S}1$ and (b) ${\cal S}2$ for a given 
c.m. energy of 500 GeV. Several interesting features are noted:
\begin{itemize}
\item Two distributions are forward-backward symmetric, which is
      due to the Majorana property of the neutralinos. This symmetry
      property can be traced back to the fact that the quartic charge
      $Q^{ij}_3$ is directly proportional to $\cos\Theta$ and the quartic 
      charges $Q^{ij}_{1,2}$ are forward--backward symmetric due to the 
      Majorana relation
      $|Q^{ij}_{\alpha L,R}(\Theta)|=|Q^{ij}_{\alpha R,L}(\pi-\Theta)|$
      where $\alpha=L,R$ stands for the electron chirality. 
\item The cross sections in the scenario ${\cal S}2$ with small selectron
      masses are much larger in size that those in the scenario ${\cal S}1$
      with very large selectron masses. This reflects the fact that the 
      $t$-- and $u$--channel selectron exchanges become dominant for small 
      selectron masses so that the production cross sections are very 
      much enhanced. One additional crucial reason for the enhancement is 
      that two neutralino states are more gaugino--dominated in the scenario
      ${\cal S}2$ than in the scenario ${\cal S}1$.
\item The cross sections due to the selectron exchanges are smaller in 
      the forward--backward directions in contradiction with
      a naive expectation of forward-backward peaking phenomena
      due to $t$-- and/or $u$--exchanges. The reason is that while the
      $Q^{ij}_3$ contribution is suppressed in the scenario ${\cal S}1$
      it is comparable in size with the other contributions with 
      opposite sign in the scenario ${\cal S}2$. Therefore, in the forward 
      and backward directions, there exist a large cancellation among 
      separate contributions.       
\item The production cross section is more sensitive to the CP phases
      in the scenario ${\cal S}2$ than in the scenario ${\cal S}1$;    
      the absolute production rate is much larger and the change due to
      different phases is larger as well in the scenario ${\cal S}2$.
      So, we can expect in the scenario ${\cal S}2$ that the cross 
      section itself can allow for a good determination of the phases. 
\end{itemize}

On the other hand the contours of the total production cross sections for 
the associated production 
of neutralinos $e^+e^-\rightarrow\tilde{\chi}^0_1\tilde{\chi}^0_2$
are displayed in Figure~7  on the
plane of the phases $\{\Phi_\mu,\Phi_1\}$ for a given energy of 500 GeV
and for two SUSY parameter sets of the scenarios
(a) ${\cal S}1$ and (b) ${\cal S}2$. 
The total production cross section is of the order of 1 fb in the
scenario ${\cal S}1$ while it is of the order of 100 fb in the scenario
${\cal S}2$. 
In the scenario ${\cal S}1$, the cross section is large when the phase 
$\Phi_1$ is around $\pi$ and the phase $\Phi_\mu$ is around $\pi/2$ 
and $3\pi/2$. However, the cross section in the scenario ${\cal S}2$
increases as the phases $\Phi_\mu$ and $\Phi_1$ approach the central point
$\{\pi,\pi\}$ along the diagonal as well as off--diagonal lines.

\subsection{Initial beam polarizations}
\label{sec:electron polarizations}

\subsubsection{Spin--spin correlations}

At future $e^+e^-$ colliders, it is expected that highly longitudinally
polarized electron and/or positron beams are available. On the other hand,
it is uncertain if high transversely polarized beams can be easily 
obtained unlike conventional $e^+e^-$ circular colliders. Nevertheless,
it is interesting to investigate the effects of the longitudinal 
and transverse polarizations of the initial beams for the determinations of
the fundamental SUSY parameters. So, in this section, we introduce a 
general formalism to describe the polarization effects of the initial beams
for any production process. Here, 
the extremely small electron and positron masses 
($m_e=5.1\times 10^{-4}\,{\rm GeV}$ ) compared to the $e^+e^-$
collision energy of the order of 100 GeV allows us to have a very much 
simplified formalism. Neglecting the electron mass renders the positron 
helicity opposite to the electron helicity in any theory preserving 
electronic chirality as shown before. Let us consider the 
neutralino--pair production process $e^+e^-\rightarrow
\tilde{\chi}^0_i\tilde{\chi}^0_j$, which is nothing but the process
under consideration. For the sake of convenience we introduce a bracket 
notation for all the helicity amplitudes $M_{\sigma\bar{\sigma}:\lambda_i
\lambda_j}=\delta_{\sigma,-\bar{\sigma}}\langle\sigma:\lambda_i\lambda_j
\rangle$ which is guaranteed by the electronic chirality invariance
and which enables us to obtain a simple form of the polarization--weighted 
squared matrix element as
\begin{eqnarray}
\Sigma_{ij}&=&\frac{1}{4}(1-P_L\bar{P}_L)\sum_{\lambda_i\lambda_j}
    \left[|\langle +:\lambda_i\lambda_j\rangle|^2
         +|\langle -:\lambda_i\lambda_j\rangle|^2\right]\nonumber\\
              &&+\frac{(P_L-\bar{P}_L)}{4}\sum_{\lambda_i\lambda_j}
    \left[|\langle +:\lambda_i\lambda_j\rangle|^2
         -|\langle -:\lambda_i\lambda_j\rangle|^2\right]\nonumber\\
              &&+\frac{P_T\bar{P}_T}{2}\cos(\alpha+\bar{\alpha})
                 \sum_{\lambda_i\lambda_j}
     \,{\cal R}\left[\langle +:\lambda_i\lambda_j\rangle
                     \langle -:\lambda_i\lambda_j\rangle^*\right]\nonumber\\
              &&+\frac{P_T\bar{P}_T}{2}\sin(\alpha+\bar{\alpha})
                 \sum_{\lambda_i\lambda_j}
     \,{\cal I}\left[\langle +:\lambda_i\lambda_j\rangle
                     \langle -:\lambda_i\lambda_j\rangle^*\right]\,,
\label{eq:squared matrix element 1}
\end{eqnarray}
where $P_L(\bar{P}_L)$ and $P_T(\bar{P}_T)$ denote the degree of 
longitudinal and transverse polarization of the electron (positron), and
$\alpha (\bar{\alpha})$ the direction of each transverse polarization with
respect to a given reference plane, for which the scattering plane is
chosen in most cases. The pictorial description of the azimuthal 
angles is given in Figure~8. We emphasize that the formalism can be applied 
to any $e^+e^-$ collision process preserving electronic chirality with 
an apporiate choice of reference 
frame to define the azimuthal angle parameters $\alpha$ and $\bar{\alpha}$.    

\vspace*{1cm} 
\begin{center}
\begin{picture}(330,200)(0,0)

\SetWidth{1.5}
\Text(10,100)[]{$e^-$}
\LongArrow(20,100)(146,100)
\LongArrow(280,100)(154,100)
\Vertex(150,100){2}
\Text(290,100)[]{$e^+$}
\DashLine(150,105)(150,180){2}
\Text(150,190)[]{$x$}
\DashLine(300,100)(320,100){2}
\Text(330,100)[]{$z$}
\SetWidth{2}
\LongArrow(80,100)(30,150)
\LongArrow(240,100)(190,50)
\SetWidth{1}
\ArrowArc(80,100)(30,90,135)
\DashLine(80,100)(80,150){2}
\ArrowArc(240,100)(30,90,225)
\DashLine(240,100)(240,150){2}
\Text(63,135)[]{$\alpha$}
\Text(210,130)[]{$\bar{\alpha}$}
\Text(25,160)[]{$P_T$}
\Text(190,40)[]{$\bar{P}_T$}

\end{picture}
\end{center}
\vskip -1cm
\noindent
Figure~8: {\it Configuration of transverse polarization vectors in the 
               center--of--mass frame. The azimuthal angles $\alpha$
               and $\bar{\alpha}$ denote the orientation of the polarization
               vectors with respect to a reference plane, for which the
               scattering plane is chosen.}

\vskip 0.5cm

The longitudinal and transverse polarizations and the azimuthal
angles for the transverse polarizations are related under CP 
transformations as follows:
\begin{eqnarray}
P_L           & \longleftrightarrow & -\bar{P}_L\,, \nonumber\\
P_T\cos\alpha & \longleftrightarrow &  \bar{P}_T\cos\bar{\alpha}\,, \nonumber\\
P_T\sin\alpha & \longleftrightarrow &  \bar{P}_T\sin\bar{\alpha},
\end{eqnarray}
while the production helicity amplitudes are related under CP transformations 
as:
\begin{eqnarray}
\langle\sigma:\lambda_i\lambda_j\rangle_{ij} \longleftrightarrow  
        \langle\sigma:-\lambda_j,-\lambda_i\rangle_{ji}\,,
\end{eqnarray}
where the subscript $ij$ means for the production of a particle
$\tilde{\chi}^0_i$ and an anti--particle $\tilde{\chi}^0_j$ according
to our spinor conventions. Denoting $\bar{\Sigma}_{ji}$ as the CP-conjugate 
polarization--correlated distribution of $\Sigma_{ij}$, one can construct 
a CP--even distribution $\frac{1}{2}(\Sigma_{ij}+\bar{\Sigma}_{ji})$ and a 
CP-odd distribution 
$\frac{1}{2}(\Sigma_{ij}-\bar{\Sigma}_{ji})$. If no CP--preserving
phases are involved or any of them are negligible in a given process,
one can find that the CP--odd distribution is proportional to the
last term of the distribution in Eq.~(\ref{eq:squared matrix element 1}),
while the CP--even distribution is composed of the other three terms.

\subsubsection{Two additional CP--even observables}

Neglecting the $Z$--boson width ($\Gamma_Z\approx 2.5\,{\rm GeV}$), which is
very small compared to the collision energies under consideration, the 
distribution (\ref{eq:squared matrix element 1}) provides us with
three CP-even observables; one of them is the unpolarized part which
has been discussed before and the other
two terms can be extracted by taking an appropriate polarization correlation.  
Longitudinal polarization yield  a differential left-right (LR) asymmetry 
$A_{LR}$ 
\begin{eqnarray}
A_{LR}=\frac{1}{4{\cal N}}\sum_{\lambda_i\lambda_j}
           \left[|\langle +;\lambda_i\lambda_j\rangle|^2
                -|\langle -;\lambda_i\lambda_j\rangle|^2\right]\,,
\end{eqnarray}
with the normalization corresponding to the unpolarized part
\begin{eqnarray}
{\cal N}=\frac{1}{4}\sum_{\lambda_i\lambda_j}
           \left[|\langle +;\lambda_i\lambda_j\rangle|^2
                +|\langle -;\lambda_i\lambda_j\rangle|^2\right]\,.
\end{eqnarray}
The LR asymmetry can readily be expressed in terms of the quartic
charges,
\begin{eqnarray}
A_{LR}&=&\bigg\{[4-(\eta_+-\eta_-)^2
                   +(\eta_++\eta_-)^2\cos^2\Theta]{Q'}^{ij}_1\nonumber\\
      &&+4\sqrt{(1-\eta^2_+)(1-\eta^2_-)}{Q'}^{ij}_2
             +4(\eta_++\eta_-)\cos\Theta {Q'}^{ij}_3\bigg\}/{\cal N}\,,
\end{eqnarray}
with, correspondingly, the expression for the normalization
\begin{eqnarray}
{\cal N}&=&[4-(\eta_+-\eta_-)^2+(\eta_++\eta_-)^2\cos^2\Theta]Q^{ij}_1
           \nonumber\\
        &&+4\sqrt{(1-\eta^2_+)(1-\eta^2_-)}Q^{ij}_2
            +4(\eta_++\eta_-)\cos\Theta Q^{ij}_3\,.
\end{eqnarray}
It will be straightforward to extract the integrated left--right asymmetry 
experimentally with the expectation that highly longitudinally polarized 
beams are available at future $e^+e^-$ linear colliders. Certainly, the 
extraction efficiency depends linearly on the degree of electron and positron 
polarization obtainable at the $e^+e^-$ collisions.

The other transverse--polarization dependent term can be separated by allowing
transverse polarization and setting longitudinal polarization to zero.
Note that the transverse term is dependent on the sum of two azimuthal angles
$\alpha$ and $\bar{\alpha}$, which must be sorted out by using 
a weight function $\sqrt{2}\cos(\alpha+\bar{\alpha})$. This projection 
requires that the scattering plane is experimentally determined event 
by event. First of all, the neutralino masses $m_{\tilde{\chi}^0_2}$ and 
$m_{\tilde{\chi}^0_1}$ are expected to be measured with good precision
through identifying the minimal and maximal values for the lepton 
invariant mass in the leptonic decay $\tilde{\chi}^0_2\rightarrow
\tilde{\chi}^0_1\ell^+\ell^-$. 
The determined masses and the four--momentum of two final
leptons enable us to determine only the polar angle between 
the $\tilde{\chi}^0_2$ flight direction and the $\tilde{\chi}^0_1$
flight direction in the laboratory frame. Certainly, if the c.m. energy is
so large that neutralino masses are negligible, the neutralino direction
can be identified with the direction of the two--lepton momentum.
However, for a moderate c.m. energy, it is not possible to completely
determine the $\tilde{\chi}^0_2$ scattering angle. Nevertheless, this 
distribution will affect the final two--lepton distribution partially so
that it is not useless to investigate the dependence of the observable
on the SUSY parameters. 
The CP--even observable ${\cal P}_T$ obtained through the angular 
projection procedure is given by
\begin{eqnarray}
{\cal P}_T\equiv\frac{1}{\sqrt{2}{\cal N}}\sum_{\lambda_i\lambda_j}\,\,
           {\cal R}\left[\langle +;\lambda_i\lambda_j\rangle
                         \langle -;\lambda_i\lambda_j\rangle^*\right]
            =-2\sqrt{2}\lambda\frac{Q^{ij}_5}{\cal N}\sin^2\Theta\,.
\end{eqnarray}
The upper figures in Figure~9 exhibit the LR asymmetries $A_{LR}$ 
and the lower ones the CP--even observables ${\cal P}_T$ for the
production of the associated pair $\tilde{\chi}^0_2$ and 
$\tilde{\chi}^0_1$ as a function of the scattering angle $\Theta$ at 
a c.m. energy of 500 GeV in the scenarios (a) ${\cal S}1$ and (b) ${\cal S}2$ 
for five combinations of the CP phases $\{\Phi_\mu,\Phi_1\}$.
We note that these two observables are also more sensitive to the CP phases
in the scenario ${\cal S}2$ than in the scenario ${\cal S}1$.
However, the sensitivity of the LR asymmetry $A_{LR}$ to the CP
phases is not strong, so that the asymmetries are not very useful 
in determining the phases. On the other hand, as discussed before, 
the CP--even observable, 
which is more sensitive to the phases, is not easy  to extract
experimentally.
Therefore, we may conclude that these two observables are not so powerful
in determining the phases in both scenarios, while satisfying the
constraints from the electron EDM measurements.

\subsubsection{One CP--odd observable}

CP violation arises in the existence of nontrivial complex couplings
in the Lagrangian. In the associated production of the neutralinos 
CP violation is reflected in the complex production amplitudes.
First of all, we find that in every diagonal production of 
neutralinos the production amplitude is purely real with the
$Z$--boson width neglected, and it leads to no CP-violation. 

Like the  CP-even observable ${\cal P}_T$, the only T-odd term 
${\cal P}_N$, which is CP-odd in the absence of any CP-even 
rescattering phases like the $Z$--boson width, can be separated by 
allowing transverse polarization
and  setting longitudinal polarization to zero with  
$\sqrt{2}\sin(\alpha+\bar{\alpha})$ as a projection angular function. 
As a result, one can obtain a CP-odd observable ${\cal P}_N$ as
\begin{eqnarray}
{\cal P}_N\equiv\frac{1}{\sqrt{2}{\cal N}}\sum_{\lambda_i\lambda_j}\,\,
           {\cal I}\left[\langle +;\lambda_i\lambda_j\rangle
                         \langle -;\lambda_i\lambda_j\rangle^*\right]
              =-2\sqrt{2}
                  \lambda\,\frac{{Q'}^{ij}_6}{\cal N}\, \sin^2\Theta\,.
\end{eqnarray}
One can check with the definition of the quartic charge ${Q'}^{ij}_6$
that if the $Z$--boson width is neglected, the T-odd observable 
${\cal P}_N$ may be non-zero only for $i\neq j$ since for $i=j$ all
the bilinear charges are real.

In CP--noninvariant theories the quartic charge ${Q'}^{ij}_6$, which 
is non--vanishing for $i\neq j$, can be expressed in terms of 
two Jarlskog-type CP-odd rephasing invariants \cite{J} of the diagonalization 
matrix $N$. In order to elaborate on this point further, we present 
the explicit form of the quartic charge; 
assuming a real $Z$-boson propagator the quartic charge
${Q'}^{ij}_6$ is given by
\begin{eqnarray}
{Q'}^{ij}_6=\frac{D_Z}{2s_W^4 c_W^4}
            \bigg[s_W^2 (D_{tL}-D_{uL})\,{\cal I}({\cal Z}_{ij} g^*_{Lij})
            -(s_W^2-1/2)(D_{tR}-D_{uR})\,{\cal I}({\cal Z}_{ij}g^*_{Rij})\bigg]
            \,.
\end{eqnarray}
Two combinations of the couplings, ${\cal I}({\cal Z}_{ij}g^*_{Lij})$
and ${\cal I}({\cal Z}_{ij}g^*_{Rij})$, are rephase-invariant. 
Using the expressions for ${\cal Z}_{ij}$, $g_{Lij}$ and
$g_{Rij}$, we can rewrite the two combinations as
\begin{eqnarray}
&& {\cal I}({\cal Z}_{ij}g^*_{Rij})
   = \frac{1}{2c_W^2}\left[{\cal I}(N_{i3}N^*_{j3}N^*_{i1}N_{j1})
                     -{\cal I}(N_{i4}N^*_{j4}N^*_{i1}N_{j1})\right]\,,
                     \nonumber\\
&& {\cal I}({\cal Z}_{ij}g^*_{Lij})
   = \frac{1}{8s_W^2c_W^2}\left[{\cal I}({N}_{i3}{N}^*_{j3}
                                         {N'} ^*_ {i2}{N'}_{j2})
                     -{\cal I}({N}_{i4}{N}^*_{j4}{N}^*_{i2}{N'}_{j2})\right]
                     \,,
\end{eqnarray}
where the primed matrix elements $N'_{i1}$ and $N'_{i2}$ are related with 
the diagonalization matrix elements $N_{i1}$ and $N_{i2}$ through
\begin{eqnarray}
N'_{i1} = c_W\, N_{i1} + s_W\, N_{i2},\qquad
N'_{i2} =-s_W\, N_{i1} + c_W\, N_{i2}\,.
\end{eqnarray}
From the expressions (37) and (38), we can draw the following consequences:
\begin{itemize}
\item Both ${\cal I}({\cal Z}_{ij}g^*_{Rij})$ and ${\cal I}({\cal Z}_{ij}
      g^*_{Lij})$ require the existence of gaugino and higgsino components 
      and a different magnitude of two higgsino components of the 
      $\tilde{\chi}^0_i$ and $\tilde{\chi}^0_j$ states. 
      This latter requirement means that $\tan\beta$ should be different 
      from unity.
\item The distribution is forward--backward asymmetric, because the angular
      dependence is determined by the difference $D_{tL,R}-D_{uL.R}$.
\item Due to the large suppression in the $t$-- and $u$--channel
      selectron exchanges, the T--odd asymmetry is very small in the
      scenario ${\cal S}1$. Moreover, the asymmetry ${\cal P}_N$ is very 
      small in the scenario ${\cal S}2$ as well. This suppression is
      because the observable requires a sizable mixing between gaugino and
      higgsino states, but the mixing is very small due to the large
      value of $|\mu|$ 
      compared to the gaugino masses $M_2$ and $|M_1|$.
\end{itemize}
As a whole, these features lead to the conclusion that the T--odd observable 
${\cal P}_N$ is not useful in measuring the CP phases directly
in both scenarios ${\cal S}1$ and ${\cal S}2$ suggested by the analysis
for the electron EDM constraints.

\subsubsection{Neutralino polarization vector}

Neutralinos are spin--1/2 particles and their polarization can be measured 
through their decays. Before we investigate the possible neutralino decays
in detail, in this section we study chargino polarization directly in the 
production process $e^+e^-\rightarrow\tilde{\chi}^0_2\tilde{\chi}^0_1$
with unpolarized initial beams.  
The polarization vector $\vec{\cal P}^{ij}=({\cal P}^{ij}_L,{\cal P}^{ij}_T,
{\cal P}^{ij}_N)$ of the produced neutralino $\tilde{\chi}^0_i$ is defined 
in the rest frame in which the axis $\hat{z}\| L$ is in the flight direction 
of $\tilde{\chi}^0_i$, $\hat{x}\| T$ rotated counter--clockwise 
in the production plane, and $\hat{y}=\hat{z}\times\hat{x}\| N$
of the decaying neutralino $\tilde{\chi}^0_i$. 
Accordingly, the component ${\cal P}^{ij}_L$ denotes the component 
parallel to the $\tilde{\chi}^0_i$ flight direction in the c.m. frame,
${\cal P}^{ij}_T$ the transverse component in the production plane, and 
${\cal P}^{ij}_N$ the component normal to the production plane. 
These three polarization components can be expressed by helicity amplitudes 
in the following way:
\begin{eqnarray}
&& {\cal P}^{ij}_L=\frac{1}{4}\sum_{\sigma=\pm}\left\{
              |\langle\sigma;++\rangle|^2+|\langle\sigma;+-\rangle|^2
             -|\langle\sigma;-+\rangle|^2-|\langle\sigma;--\rangle|^2
                                           \right\}/{\cal N}
              \,,\nonumber\\
&& {\cal P}^{ij}_T=\frac{1}{2}{\cal R}\bigg\{\sum_{\sigma=\pm}\left[
              \langle\sigma;++\rangle\langle\sigma;-+\rangle^*
             +\langle\sigma;--\rangle\langle\sigma;+-\rangle^*
                          \right]\bigg\}/{\cal N}\,,\nonumber\\
&& {\cal P}^{ij}_N=\frac{1}{2}{\cal I}\bigg\{\sum_{\sigma=\pm}\left[
              \langle\sigma;--\rangle\langle\sigma;+-\rangle^*
             -\langle\sigma;++\rangle\langle\sigma;-+\rangle^*
                           \right]\bigg\}/{\cal N}\,.
\end{eqnarray}
The longitudinal, transverse and normal components of the 
$\tilde{\chi}^0_i$ polarization vector can be easily obtained from the 
production helicity amplitudes. Expressed in terms of the quartic charges, 
they read:
\begin{eqnarray}
&& {\cal P}^{ij}_L = 4 \bigg\{2(1-\mu^2_i-\mu^2_j)\cos\Theta Q'^{ij}_1
       +4\mu_i\mu_j\cos\Theta Q'^{ij}_2\nonumber\\
&& { } \hskip 3cm +\lambda^{1/2}[1+\cos^2\Theta
         -(\mu^2_i-\mu^2_j)\sin^2\Theta]Q'^{ij}_3\bigg\}/{\cal N}
         \,,\nonumber\\                          
&& {\cal P}^{ij}_T = -8\bigg\{ [(1-\mu^2_i+\mu^2_j) Q'^{ij}_1
       +\lambda^{1/2}Q'^{ij}_3\cos\Theta]\mu_i
       +(1+\mu^2_i-\mu^2_j)\mu_j Q'^{ij}_2\bigg\}\sin\Theta/{\cal N}
         \,,\nonumber\\
&& {\cal P}^{ij}_N = 8\lambda^{1/2}\mu_j\sin\Theta Q_4^{ij}/{\cal N}\,,
\label{eq:polarization vector}
\end{eqnarray}
where the reduced masses $\mu^2_i=m^2_{\tilde{\chi}^\pm_i}/s$.
The longitudinal and transverse components are P--odd and CP--even, and 
the normal component is P--even and CP--odd. 

The normal polarization component can only be generated by complex 
production amplitudes. Non--zero phases are present in the fundamental 
SUSY parameters if CP is broken in the supersymmetric interaction. 
Also the non--zero width of the $Z$ boson and loop corrections generate 
non--trivial phases; 
however, the width effect is negligible for high energies as
mentioned before, and the effects due to radiative corrections are small 
as well. So, the normal component is effectively generated by the complex 
SUSY couplings. 
As the selectron--exchange contributions can be ignored in the scenario
${\cal S}1$, the CP--odd quartic charge $Q^{ij}_4$ simplifies to
\begin{eqnarray}
Q^{ij}_4 =\frac{|D_Z|^2}{c^4_W s^4_W}\,\left(s^2_W-\frac{1}{4}\right)\,
          {\cal I}\left(Z^2_{ij}\right)\,.
\end{eqnarray}
Since the presently measured value of $s^2_W$ is $0.2315$ \cite{PDG98} very 
close to $0.25$, the quartic charge $Q^{ij}_4$ is extremely suppressed in the 
scenario ${\cal S}1$. However, in the scenario 
${\cal S}2$, the quartic charge can be relatively large without such a
big suppression as shown in Figure~10. In this case, the normal
polarization is of the order of 10 \%, which is really sizable
for non--trivial CP phases and very sensitive to the CP phases 
$\{\Phi_\mu,\Phi_1\}$. So, it is expected to give stringent constraints 
on the phases $\Phi_\mu$. These strong constraints will be explicitly
demonstrated in the following.

\section{Neutralino Decays}

\subsection{Decay density matrix}

Assuming the lightest neutralino $\tilde{\chi}^0_1$ to be the
lightest supersymmetric particle (LSP), several mechanisms contribute to 
the leptonic decays of the neutralino $\tilde{\chi}^0_i$ ($i\geq 2$):
\begin{eqnarray*}
\tilde{\chi}^0_i(q_i) \rightarrow\tilde{\chi}^0_1 (q_0) +\ell^-
(q)+ \ell^+(\bar{q})
\end{eqnarray*}
In particular, the leptonic three body decay of the second lightest 
neutralino, $\tilde{\chi}^0_2\rightarrow\tilde{\chi}^0_1\ell^+\ell^-$, is 
known to be very important because the end point of the lepton
invariant mass distribution gives us direct information on
the mass difference between $\tilde{\chi}^0_2$ and $\tilde{\chi}^0_1$,
which provides us with a stringent constraints on MSSM parameters.

Although we will take into account only electrons and muons for the
final state leptons, let us make some comments on the other possible
leptonic decay of the second lightest neutralino, $\tilde{\chi}^0_2
\rightarrow\tilde{\chi}^0_1\tau^+\tau^-$ \cite{BCDP}. Since $\tau$ is
the heaviest lepton with a much larger mass ($1.777$ GeV) than the
other leptons and it couples with Higgs bosons with the strength 
proportional to $\tan\beta$, the branching fraction of this
leptonic decay mode can be very different depending on the value
of $\tan\beta$ and the Higgs mass spectrum. Actually, the mode
is known to be very much enhanced due to the Higgs exchanges at
large $\tan\beta$ \cite{BCDP}. Furthermore, the polarization of $\tau$ 
can be 
observed through the decay distributions, which strongly depend on the 
parent $\tau$ polarization \cite{THP}. 

Due to the missing tau neutrinos, one would not be able to
measure the invariant mass of two tau leptons experimentally. 
Nevertheless the $\tau$ polarization or the invariant mass of the two
$\tau$ jets might be seen in future collider experiments.
We note that in $\tau\rightarrow\rho$ or $a_1$ decays, the final
vector meson carries a substantial part of the parent $\tau$ momentum, 
therefore the smearing of the distribution is less severe than
for decays into $\pi^\pm$, $\mu$ and $e$.
Let us give several comments on the tau decay mode.
First of all, for $\tilde{\tau}$, the effects of the Yukawa couplings 
and slepton left--right mixing could be very important for large $\tan\beta$.
Their leading contribution flips the chirality
of the $\tau$ lepton \cite{Nojiri}. 
Secondly, for the three body decays, studying 
the correlation of two tau decay distributions would reveal the helicity 
flipping and conserving contributions separately. Thirdly,
staus could be lighter than the other sleptons for various 
reseaons. The running of stau soft SUSY breaking masses from the
Planck scale and stau left--right mixing could enhance decays
into $\tilde{\chi}^0_1\tau^+\tau^-$. Experimental consequences
of such scenarios have recently been widely discussed.
Also models with lighter third generation sparticles
have been naturally constructed without causing the
flavor changing neutral current problem. The three body decay 
branching ratio and the decay
distribution might be different from those for leptons in the first
two generations. Since the study of the decay $\tilde{\chi}^0_2\rightarrow
\tilde{\chi}^0_1\tau^+\tau^-$ in addition to the other
leptonic modes could be an important handle to identify
such models, we plan to present a detailed investigation about all these 
interesting features in the near future.  
\vspace*{-1cm}
\begin{center}
\begin{picture}(350,220)(0,0)
\Text(0,120)[]{$\tilde{\chi}^0_i$}
\ArrowLine(10,120)(40,120)
\Photon(10,120)(40,120){3}{7}
\ArrowLine(40,120)(70,150)
\Photon(40,120)(70,150){3}{7}
\Text(79,150)[]{$\tilde{\chi}^0_1$}
\Photon(40,120)(60,90){4}{8}
\Text(30,95)[]{$Z$}
\ArrowLine(60,90)(90,90)
\Text(100,90)[]{$\ell^-$}
\ArrowLine(80,60)(60,90)
\Text(85,55)[]{$\ell^+$}
\Text(125,120)[]{$\tilde{\chi}^0_i$}
\ArrowLine(135,120)(165,120)
\Photon(135,120)(165,120){3}{7}
\ArrowLine(165,120)(195,150)
\Text(200,150)[]{$\ell^-$}
\Line(185,90)(165,120)
\Line(188,90)(167,122)
\Text(165,95)[]{$\tilde{\ell}_{L,R}$}
\ArrowLine(185,90)(215,90)
\Photon(185,90)(215,90){3}{7}
\Text(225,90)[]{$\tilde{\chi}^0_1$}
\ArrowLine(205,60)(185,90)
\Text(210,55)[]{$\ell^+$}
\Text(250,120)[]{$\tilde{\chi}^0_i$}
\ArrowLine(260,120)(290,120)
\Photon(260,120)(290,120){3}{7}
\ArrowLine(320,150)(290,120)
\Text(325,150)[]{$\ell^+$}
\Line(290,120)(310,90)
\Line(292,122)(313,90)
\Text(290,95)[]{$\tilde{\ell}_{L,R}$}
\ArrowLine(310,90)(340,90)
\Photon(310,90)(340,90){3}{7}
\Text(350,90)[]{$\tilde{\chi}^0_1$}
\ArrowLine(310,90)(335,55)
\Text(345,55)[]{$\ell^-$}
\end{picture}\\
\end{center}
\vspace*{-0.6cm}
\noindent
Figure~11: {\it Neutralino decay mechanisms; the exchange of the neutral Higgs 
               boson is neglected because of the tiny electron Yukawa
               coupling.}
\bigskip

The diagrams contributing to the process $\tilde{\chi}^0_i\rightarrow
\tilde{\chi}^0_j\ell^+\ell^-$ with $\ell=e$, $\mu$ are shown in Fig.~11 
for the decay into lepton pairs. Here, the exchange of  the neutral 
Higgs boson [replacing the $Z$ boson] are neglected since the couplings 
to the light first and second generation SM leptons are very small. 
In this case, all the components of the decay matrix elements 
are of the left/right current$\times$current form which, after a simple 
Fierz transformation, may be written for lepton final states as
\begin{eqnarray}
D\left(\tilde{\chi}^0_i\rightarrow\tilde{\chi}^0_1 \ell^-\ell^+\right)
 = \frac{e^2}{s'}D_{\alpha\beta}
   \left[\bar{u}(\tilde{\chi}^0_1) \gamma^\mu P_\alpha 
               u(\tilde{\chi}^0_i)\right]
   \left[\bar{u}(\ell^-)  \gamma_\mu P_\beta  v(\ell^+)\right]\,,
\label{eq:neutralino decay amplitude}
\end{eqnarray}
with the generalized bilinear charges for the decay amplitudes:
\begin{eqnarray}
D_{LL}&=&+\frac{D_Z'}{s_W^2c_W^2}(s_W^2 -\frac{1}{2}){\cal Z}_{1i}
              -D_{u'L}\,g_{L1i}\,,\nonumber\\ 
D_{LR}&=&+\frac{D_Z'}{c_W^2}{\cal Z}_{1i}
              +D_{t'R}\,g_{R1i}\,,\nonumber\\
D_{RL}&=&-\frac{D_Z'}{s_W^2c_W^2}(s_W^2 -\frac{1}{2}){\cal Z}^*_{1i}
              +D_{t'L}\,g^*_{L1i}\,,\nonumber\\
D_{RR}&=&-\frac{D_Z'}{c_W^2}{\cal Z}^*_{1i}
              -D_{u'R}\,g^*_{R1i}\,,
\end{eqnarray}
where the chiralities $\alpha/\beta$ stand for the 
$\tilde{\chi}^0_1/\ell^-$ chiralities and $s'-,t'-,$ and $u'-$channel 
propagators and the couplings ${\cal Z}_{ij},g_{Lij},$ and 
$g_{R1i}$ are given in the section for the production helicity
amplitudes. The Mandelstam variables $s',t',u'$ are
defined in terms of the 4-momenta of $\tilde{\chi}^0_1, \ell^-$ and 
$\ell^+$, respectively, as
\begin{eqnarray}
s'=(q+\bar{q})^2 \ \ , \ \ t'=(q_0+q)^2 \ \ , \ \ u'=(q_0 +\bar{q})^2\,. 
\end{eqnarray}
The decay distribution of a neutralino with polarization vector 
$n^\mu$ is
\begin{eqnarray}
|{\cal D}|^2(n) &=& 
-4(t'- m_{\chi_i^0}^2)(t'- m_{\chi_1^0}^2)(D_1-D_3) 
-4(u'- m_{\chi_i^0}^2)(u'- m_{\chi_1^0}^2)(D_1+D_3)   \nonumber\\
&& -8 m_{\tilde{\chi}^0_i} m_{\tilde{\chi}^0_1} s'D_2 \nonumber \\
&& -8 (n\cdot\bar{q})\left[
 m_{\tilde{\chi}^0_i}(m_{\tilde{\chi}^0_1}^2-u')(D'_1+D'_3)
+m_{\tilde{\chi}^0_1}(m_{\tilde{\chi}^0_i}^2-t')D'_2\right]  \nonumber\\
&& +8(n\cdot q)\left[
-m_{\tilde{\chi}^0_i}(m_{\tilde{\chi}^0_1}^2-t')(D'_1-D'_3)
+m_{\tilde{\chi}^0_1}(m_{\tilde{\chi}^0_i}^2-u')D'_2\right]  \nonumber\\ 
&& +16 m_{\tilde{\chi}^0_1}\langle q_i n q \bar{q}\rangle D_4\,,
\label{eq:decay distribution}
\end{eqnarray}
where $n_\mu$ is the $\tilde{\chi}^0_i$ spin 4--vector and 
$\langle q_i n q \bar{q}_2\rangle\equiv\epsilon_{\mu\nu\rho\sigma} 
q_i^{\mu}n^{\nu}q^{\rho}\bar{q}^{\sigma}$.
Here, the quartic charges $\{D_1\,{\rm to}\, D_4\}$ and 
$\{D'_1\,{\rm to}\, D'_3\}$ for the neutralino decays are defined by
\begin{eqnarray}
&& D_1=\frac{1}{4}
            \left[|D_{RR}|^2+|D_{LL}|^2
                 +|D_{RL}|^2+|D_{LR}|^2\right]\,,  \nonumber\\
&& D_2=\frac{1}{2}{\cal R}
            \left[D_{RR}D^{*}_{LR}
                 +D_{LL}D^{*}_{RL}\right]\,,       \nonumber\\
&& D_3=\frac{1}{4}
            \left[|D_{LL}|^2+|D_{RR}|^2
                 -|D_{RL}|^2-|D_{LR}|^2\right]\,,  \nonumber\\
&& D_4=\frac{1}{2}{\cal I}
            \left[D_{RR}D^{*}_{LR}
                 +D_{LL}D^{*}_{RL}\right]\,,       \nonumber\\
&& D'_1=\frac{1}{4}
             \left[|D_{RR}|^2+|D_{RL}|^2
                  -|D_{LR}|^2-|D_{LL}|^2\right]\,, \nonumber\\
&& D'_2=\frac{1}{2}{\cal R}
             \left[D_{RR}D^{*}_{LR}
                  -D_{LL}D^{*}_{RL}\right]\,,      \nonumber\\
&& D'_3=\frac{1}{4}
             \left[|D_{RR}|^2+|D_{LR}|^2
                  -|D_{RL}|^2-|D_{LL}|^2\right]\,.
\end{eqnarray}
If needed, the polarization of the final neutralino $\tilde{\chi}^0_j$
can be incorporated in a straightforward manner although the
decay distribution will be more complicated in its form.

For the subsequent discussion of the angular correlations between 
two neutralinos, it is convenient to determine the decay spin density 
matrix $\rho_{\lambda\lambda'}\sim {\cal D}_\lambda {\cal 
D}^*_{\lambda'}$. In general, the decay amplitude for a spin--1/2 particle 
and its complex conjugate can be expressed as 
\begin{eqnarray}
{\cal D}(\lambda) = \Gamma\, u(q,\lambda),\qquad 
{\cal D}^*(\lambda') = \bar{u}(q,\lambda')\,\bar{\Gamma},
\end{eqnarray}
with the general spinor structure $\Gamma$ and $\bar{\Gamma}=\gamma^0
\Gamma^\dagger$. Then we use the general formalism to calculate
the decay density matrix involving a particle with four
momentum $q$ and mass $m$ by introducing three space-like four vectors 
$n^a_\mu$ ($a=1,2,3$) which together with $q/m\equiv n^0$ form an orthonormal 
set:
\begin{eqnarray}
g^{\mu\nu}\, n^a_\mu n^b_\nu =g^{ab},\qquad 
g_{ab}\, n^a_\mu n^b_\nu = g_{\mu\nu}\,,
\end{eqnarray}
where $g^{\mu\nu}={\rm diag}(1,-1,-1,-1)$ and $g^{ab}={\rm diag}(1,-1,-1,-1)$
with $a,b=\{0\,-\,4\}$. A convenient choice for the explicit 
form of $n^a$ is in a coordinate system where the direction of the 
three--momentum of the particle is $\hat{q}=(\sin\theta,0,\cos\theta)$ 
lying on the $x$-$z$ plane:
\begin{eqnarray}
n^1=(0,\cos\theta,0,-\sin\theta)\,,\qquad
n^2=(0,0,1,0)\,,\qquad 
n^3=\frac{1}{m}\left(|\vec{q}|,E\hat{q}\right)\,.
\end{eqnarray}
Then in this reference frame, $n^{1,2,3}$ describe transverse, normal
and longitudinal polarization of the particle.

With the four--dimensional basis of normal four--vectors $\{n^0,n^1,n^2,n^3\}$,
we can derive the so--called Bouchiat--Michel formula \cite{BM}
\begin{eqnarray}
u(q,\lambda)\bar{u}(q,\lambda')=\frac{1}{2}
   \left[\delta_{\lambda\lambda'}+\gamma_5\not\!{n}^a\tau^a_{\lambda'\lambda}
   \right](\not\!{q}+m)\,,
\end{eqnarray}
which can be used to compute the squared, normalized decay density matrix 
$\rho_{\lambda\lambda'}$
\begin{eqnarray}
\rho_{\lambda\lambda'}\equiv
      \frac{{\cal D}(\lambda){\cal D}^*(\lambda')}{\sum_{\lambda}
            |{\cal D}(\lambda)|^2} = \frac{1}{2}\left[\delta_{\lambda\lambda'}
           +\frac{Y^{a}}{X}\tau^{a}_{\lambda'\lambda}\right]
\end{eqnarray}
where $\tau^a$ ($a=1,2,3$) are the Pauli matrices and
the four kinematic functions $X$ and $Y^a$ ($a=1,2,3$) 
\begin{eqnarray}
X &=& -8(t'- m_{\chi_i^0}^2)(t'- m_{\chi_1^0}^2)(D_1-D_3) 
      -8(u'- m_{\chi_i^0}^2)(u'- m_{\chi_1^0}^2)(D_1+D_3)   \nonumber\\
   && -16 m_{\tilde{\chi}^0_i} m_{\tilde{\chi}^0_1} s'D_2 \nonumber \\
Y^a &=&-16 (n^a\cdot\bar{q})\left[
        m_{\tilde{\chi}^0_i}(m_{\tilde{\chi}^0_1}^2-u')(D'_1+D'_3)
       +m_{\tilde{\chi}^0_1}(m_{\tilde{\chi}^0_i}^2-t')D'_2\right]\nonumber\\
    && +16(n^a\cdot q)\left[
       -m_{\tilde{\chi}^0_i}(m_{\tilde{\chi}^0_1}^2-t')(D'_1-D'_3)
       +m_{\tilde{\chi}^0_1}(m_{\tilde{\chi}^0_i}^2-u')D'_2\right]\nonumber\\ 
    && +32 m_{\tilde{\chi}^0_1}\langle q_i n^a q \bar{q}\rangle D_4\,,
\end{eqnarray}
with $n^a$ ($a=1,2,3$) three vectors forming the polarization basis 
for the decaying spin--1/2 particle.

\subsection{Branching ratios}

Since the reconstruction of the neutralino-pair production depends
on the efficient use of the neutralino decay modes, it is necessary to 
estimate the branching fraction of each decay mode. Since we assume that 
the lightest neutralino is the LSP and we are interested only in the decay 
of the second--lightest neutralino $\tilde{\chi}^0_2$, we can classify 
the decay modes as follows:
\begin{eqnarray}
&& \tilde{\chi}^0_2\rightarrow Z^*\tilde{\chi}^0_1, H^*\tilde{\chi}^0_1
                   \rightarrow \tilde{\chi}^0_1\ell^+\ell^-,
                               \tilde{\chi}^0_1 q\bar{q}\,,\nonumber\\
&& \tilde{\chi}^0_2\rightarrow \ell\tilde{\ell}^*,\nu\tilde{\nu}^*,q\tilde{q}^*
                   \rightarrow  \tilde{\chi}^0_1\ell^+\ell^-,
                               \tilde{\chi}^0_1 q\bar{q}\,.
\end{eqnarray}
Besides, if the mass $m_{\tilde{\chi}^\pm_1}$ is smaller than the
neutralino mass $m_{\tilde{\chi}^0_2}$, the lightest chargino 
$\tilde{\chi}^\pm_1$ can enter the neutralino decay chain
via $\tilde{\chi}^0_2\rightarrow\tilde{\chi}^\pm_1 W^{\mp *},
\tilde{\chi}^\pm_1 H^{\mp *}$. Concerning the neutralino decays, 
there are several aspects worthwhile to be commented on:
\begin{itemize}
\item For the first and second generation fermions, the Higgs--exchange
      diagrams are suppressed unless $\tan\beta$ is very large.
\item The experimental bounds on the Higgs particles are very stringent
      so that the two--body decays $\tilde{\chi}^0_2\rightarrow 
      H\tilde{\chi}^0_1$ and $\tilde{\chi}^0_2\rightarrow H^\pm
      \tilde{\chi}^\mp_1$ are expected to be not available or at least
      strongly suppressed.
\item The lightest chargino and the second--lightest neutralinos are
      almost degenerate in the gaugino--dominated parameter space  
      so that the charged decays such as $\tilde{\chi}^0_2
      \rightarrow\tilde{\chi}^\pm_1\ell^\mp\nu$ will be highly suppressed.
\end{itemize}
Nevertheless, we calculate the leptonic branching fractions
${\cal B}(\tilde{\chi}^0_2\rightarrow\tilde{\chi}^0_1\ell^+\ell^-)$
fully incorporating all the possible decay modes of the neutralino
$\tilde{\chi}^0_2$ while neglecting the Higgs-exchange contributions
for a small $\tan\beta=3$. In our numerical analysis, we assume 200
GeV for a common soft--breaking slepton mass and 500 GeV for 
a common soft--breaking squrk mass in the scenario ${\cal S}2$ while
we take 10 TeV for all the soft--breaking sfermion masses in
the scenario ${\cal S}1$.
Figure~12 shows ${\cal B}(\tilde{\chi}^0_2
\rightarrow\tilde{\chi}^0\ell^+\ell^-)$ for $\ell=e$ or $\mu$ in
the scenarios (a) ${\cal S}1$ and (b) ${\cal S}2$. We find that
the branching fractions are very sensitive to $\Phi_1$ only around 
$\Phi_\mu=0,2\pi$ in the scenario ${\cal S}1$, while it depends very 
strongly on $\Phi_1$ and $\Phi_\mu$ on (almost) the whole space of the 
phases
in the scenario ${\cal S}2$. Furthermore, the branching fraction 
${\cal B}(\tilde{\chi}^0_2\rightarrow \tilde{\chi}^0_1\ell^+\ell^-)$ is 
very much enhanced in the scenario ${\cal S}2$ 
because the slepton--exchange contributions due to mainly the gaugino
components of the neutralinos become dominant due to the small slepton
masses while the large value of $|\mu|$ suppressing the higgsino
components. Consequently, the branching ratio of the leptonic
decay of the second lightest neutralino $\tilde{\chi}^0_2\rightarrow
\tilde{\chi}^0_1\ell^+\ell^-$ is very sensitive to the values of
the underlying parameters, in particular, the CP phases.

\section{Spin and Angular correlations}
\label{sec:correlation}

\subsection{Correlations between production and decay}

In this section, we provide a general formalism to describe the
spin correlations between production and decay for the
process $e^+e^-\rightarrow\tilde{\chi}^0_i\tilde{\chi}^0_j$
followed by the sequential leptonic decay $\tilde{\chi}^0_i\rightarrow
\tilde{\chi}^0_1\ell^-\ell^+$. For the sake of convenience,
we do not consider the initial beam polarization, which can  however be easily
implemented. Formally,  the spin--correlated distribution is obtained by 
taking the following sum over the helicity indices of the intermediate 
neutralino state $\tilde{\chi}^0_i$
by folding the decay density matrix and the production matrix formed 
with production helicity amplitudes:  
\begin{eqnarray}
\sum_{\rm corr} \equiv \pi^2 \alpha^2 
                      \sum_{\lambda\lambda'}\sum_{\bar{\lambda}}\sum_\sigma
                      \langle\sigma:\lambda\bar{\lambda}\rangle
                      \langle\sigma:\lambda'\bar{\lambda}\rangle^*
                      \rho_{\lambda\lambda'}
                 = 2\pi^2\alpha^2 \left[\Sigma_{\rm unp}
                     +\frac{Y_3}{X}{\cal P}+\frac{Y_1}{X}{\cal V}
                     +\frac{Y_2}{X}\bar{\cal V}\right] 
\end{eqnarray}
where the functions of the scattering angle $\Theta$ are given in terms
of the production helcity amplitudes by
\begin{eqnarray}
\Sigma_{\rm unp}&=&\frac{1}{4}\sum_{\sigma=\pm}
      \bigg[|\langle\sigma;++\rangle|^2+|\langle\sigma;+-\rangle|^2
           +|\langle\sigma;-+\rangle|^2+|\langle\sigma;--\rangle|^2
      \bigg],\nonumber \\
{\cal P}&=&\frac{1}{4}\sum_{\sigma=\pm}
      \bigg[|\langle\sigma;++\rangle|^2+|\langle\sigma;+-\rangle|^2
           -|\langle\sigma;-+\rangle|^2-|\langle\sigma;--\rangle|^2
      \bigg], \nonumber\\
{\cal V}&=&\frac{1}{2}\sum_{\sigma=\pm}
      {\cal R}\bigg\{\langle\sigma;-+\rangle\langle\sigma;++\rangle^*
                    +\langle\sigma;--\rangle\langle\sigma;+-\rangle^*\bigg\},
             \nonumber \\
\bar{\cal V}&=&\frac{1}{2}\sum_{\sigma=\pm}
      {\cal I}\bigg\{\langle\sigma;-+\rangle\langle\sigma;++\rangle^*
                    +\langle\sigma;--\rangle\langle\sigma;+-\rangle^*\bigg\}.
\label{eq:polarization component}
\end{eqnarray}
Notice that the above combinations are directly related with the 
polarization vector of the neutralino $\tilde{\chi}^0_2$ as
follows
\begin{eqnarray}
{\cal P}^{i1}_L=\frac{{\cal P}}{\Sigma_{\rm unp}}\,,\ \
{\cal P}^{i1}_T=\frac{{\cal V}}{\Sigma_{\rm unp}}\,,\ \
{\cal P}^{i1}_N=\frac{\bar{\cal V}}{\Sigma_{\rm unp}} \,.
\end{eqnarray}

Combining production and decay, we obtain the fully--correlated 6-fold 
differential cross section
\begin{eqnarray}
\frac{{\rm d}\sigma}{{\rm d}\Phi}=\frac{\pi\alpha^2\beta}{8s}\, \Sigma_{\rm unp}
\,{\cal B}(\tilde{\chi}^0_i\rightarrow\tilde{\chi}^0_1\ell^-\ell^+)
\,\bigg[1+P_z{\cal P}^{i1}_L+P_x{\cal P}^{i1}_T+P_y{\cal P}^{i1}_N\bigg],
\label{eq:total differential c.s.}
\end{eqnarray}
where 
\begin{eqnarray}
P_x=\frac{Y_1}{X}\,, \ \
P_y=\frac{Y_2}{X}\,, \ \
P_z=\frac{Y_3}{X}\,,
\end{eqnarray}
with the phase space volume element ${\rm d}\Phi= {\rm d}\cos\Theta
{\rm d}x_1{\rm d}x_2{\rm d}\cos\theta_1{\rm d}\phi_1{\rm d}\phi_{12}$.
The angular variable $\theta_1$ is the polar angle of the $\ell^-$ in the 
$\tilde{\chi}^0_i$ rest frame with respect to the original 
flight direction in the laboratory frame, and $\phi_1$ the 
corresponding azimuthal angle with respect to the production plane, and 
$\phi_{12}$ is the relative azimuthal angle of $\ell^+$ 
along the $\ell^-$ direction with respect to the production plane.
On the other hand, the opening angle $\theta_{12}$ between
the $\ell^-$ and $\ell^+$ is fixed once the lepton energies are
known. The pictorial description of the kinematical variables 
for the decay $\tilde{\chi}^0_2\rightarrow\tilde{\chi}^0_1\ell^+\ell^-$
is presented
in Fig.~13. The dimensionless parameters $x_1$ and $x_2$ denote the
lepton energy fractions
\begin{eqnarray}
E_{\ell^-} = \frac{m_{\tilde{\chi}^0_2}}{2}x_1,\qquad
E_{\ell^+} = \frac{m_{\tilde{\chi}^0_2}}{2}x_2,
\end{eqnarray}
with respect to the neutralino mass $m_{\tilde{\chi}^0_i}$ divided
by a factor of two. The kinematically--allowed range for the variables
is determined by the kinematic conditions
\begin{eqnarray}
&& 0\leq \Theta    \leq \pi\,, \nonumber\\
&& 0\leq \theta_1  \leq \pi\,, \ \
   0\leq \phi_1    \leq 2\pi\,,\ \
   0\leq \phi_{12} \leq 2\pi\,,\nonumber\\
&& 0\leq x_{1,2}   \leq 1-r_{21}\,,   \ \
   (1-x_1)(1-x_2)\geq r_{21} \,, \ \
   x_1+x_2\geq 1-r_{21}\,,
\end{eqnarray}
where $r_{21}=m^2_{\tilde{\chi}^0_1}/m^2_{\tilde{\chi}^0_2}$,
and the masses of the final--state leptons are neglected.

\vspace*{1cm} 
\begin{center}
\begin{picture}(330,250)(0,-25)

\SetWidth{2}
\Text(17,100)[]{$e^-$}
\LongArrow(25,100)(146,100)
\LongArrow(275,100)(154,100)
\LongArrowArc(150,100)(40,180,225)
\Text(100,79)[]{$\Theta$}
\Text(283,100)[]{$e^+$}
\LongArrow(153,103)(210,160)
\LongArrow(147,97)(90,40)
\Text(175,142)[]{$\tilde{\chi}^0_2$}
\Text(75,35)[]{$\tilde{\chi}^0_1$}
\SetWidth{1.5}
\Vertex(150,100){2}
\LongArrow(210,160)(210,120)
\Text(210,113)[]{$\ell^+$}
\LongArrow(210,160)(173,197)
\Text(170,185)[]{$\ell^-$}
\DashLine(210,160)(250,160){3}
\Text(265,160)[]{$\tilde{\chi}^0_1$}
\SetWidth{1}
\LongArrowArc(163.333,210)(25,0,54)
\Text(199,223)[]{$\phi_{12}$}
\DashLine(173,197)(148,224){2}
\Line(5,210)(163,210)
\DashLine(163,210)(237,210){3}
\Line(237,210)(295,210)
\Line(295,210)(295,165.55)
\DashLine(295,160.55)(295,90){3}
\Line(295,90)(295,15)
\Line(295,15)(5,15)
\Line(5,15)(5,210)
\Line(190,250)(163.333,210)
\DashLine(163.333,210)(130,160){4}
\DashLine(130,160)(265,45){4}
\Line(295,90)(325,135)
\DashLine(265,45)(295,90){4}
\Line(325,135)(190,250)
\DashLine(212,162)(240,190){4}
\LongArrowArc(210,160)(25,45,135)
\Text(210,195)[]{$\theta_1,\phi_1$}
\end{picture}
\end{center}
\vskip -1cm
\noindent
Figure~13: {\it Configuration of momenta in the $e^+e^-$ c.m. frame for the
               associated production of neutralinos and in the rest frame of
               the decaying neutralino $\tilde{\chi}^0_2$.}

\vskip 0.5cm

\subsection{Total cross section of the correlated process}

The total cross section for the correlated process 
$e^+e^-\rightarrow\tilde{\chi}^0_1\tilde{\chi}^0_2\rightarrow
\tilde{\chi}^0(\tilde{\chi}^0_1\ell^-\ell^+)$ is given by integrating 
the differential cross section (\ref{eq:total differential c.s.}) over 
the full range of 
the 6--dimensional phase space. More simply, the cross section is 
given as the multiplication of the integrated production cross section
$\sigma(e^+e^-\rightarrow\tilde{\chi}^0_2\tilde{\chi}^0_1)$ and 
the branching fraction ${\cal B}(\tilde{\chi}^0_2\rightarrow\tilde{\chi}^0_1
\ell^+\ell^-)$ because the correlation effects are washed out after
integration over the phase space volume. 

The total cross section is displayed in Figure~14 as the contour plots 
on the plane of two CP phases $\{\Phi_\mu,\Phi_1\}$ in two scenarios 
(a) ${\cal S}1$ and (b) ${\cal S}2$.
First of all, we note that the total cross section in the scenario
${\cal S}1$ is very small. By definition, the scenario ${\cal S}1$ has
extremely large selectron masses so that only the $Z$--exchange diagram 
involving only the higgsino composition of the neutralinos
contributes. However, the higgsino composition of the neutralinos is at most 
40\% and in most cases 20\% for $\tilde{\chi}^0_2$ while
it is at most 20\% and in most cases 10\% for $\tilde{\chi}^0_1$.
Therefore, it is naturally expected to have a strongly--suppressed cross 
section in the scenario. On the other hand, the total cross
section is very much enhanced in the scenario ${\cal S}2$, mainly due
to large gaugino composition and small selectron masses which 
contribute to the cross section through the $t$-- and $u$--channel
exchanges. So, we can conclude that it is possible to have
a relatively large cross section if the higgsino composition of the
neutralinos is large or the slepton masses are small \cite{EHN}.

Quantitatively the scenario ${\cal S}1$ will present a few events 
of the neutralino process for a integrated luminosity of the order of 
100 fb$^{-1}$ while the scenario ${\cal S}2$ give a few thousand events 
for the same integrated luminosity. So, a good precision measurement of 
the relevant SUSY parameters might be performed in the
scenario ${\cal S}2$ at future high luminosity $e^+e^-$ collider 
experiments such as TESLA, while it might be difficult in the 
scenario ${\cal S}1$. As can be read from the figures, 
the cross section increases as the phase $\Phi_1$ approaches $\pi$ in both 
scenarios. However, the dependence of the cross section on the phase 
$\Phi_\mu$ is very different in two scenarios. In the scenario ${\cal S}2$, 
the cross section increases monotonically as $\Phi_\mu$ approaches $\pi$, but
it becomes maximal at a non-trivial value of $\Phi_\mu$
between $0$ ($2\pi$) and $\pi$ in the scenario ${\cal S}1$.

Compared with the dependence of the neutralino masses on the CP
phases displayed in Figure~4, the cross section can be larger
for larger neutralino masses and vice versa for quite large
region of the CP phases. This implies that within the range
allowed in the scenarios the signals with larger masses can have
more possibility of being detected while those with smaller masses
may escape detection. In this sense, future high luminosity experiments
can give constraints on the CP phases simply by putting 
the upper limits on the event rates.

\subsection{Dilepton invariant mass distributions}

The invariant mass of two final--state leptons $m_{\ell\ell}$,
which is nothing but the square root of the Mandelstam variable,
$\sqrt{s'}$,
\begin{eqnarray}
m_{\ell\ell}= \sqrt{s'} =m^2_{\tilde{\chi}^0_2}
              \left(x_1+x_2-1+r_{21}\right)\,,
\end{eqnarray}
is a Lorentz--invariant kinematical variable so that it is easy to 
reconstruct by measuring the energies of two final--state leptons
\cite{NY}. 
Furthermore, the distribution for the invariant mass $m_{\ell\ell}$ 
is independent of the specific production process for the 
decaying neutralino. This factorization is due to the fact that
the invariant mass does not involve any angular variables
describing the decays so that the polarization of the decaying neutralino
is not effective. 

Figure~15 shows the two--lepton invariant mass distribution
in the scenarios (a) ${\cal S}1$
and (b) ${\cal S}2$. This distribution must reflect the two--lepton 
invariant mass distribution of the neutralino decay $\tilde{\chi}^0_2
\rightarrow\tilde{\chi}^0_1\ell^+\ell^-$ itself multiplied by the
total production cross section $\sigma(e^+e^-\rightarrow\tilde{\chi}^0_2
\tilde{\chi}^0_1)$.  As shown before, the leptonic branching
ratio is very sensitive to the values of the underlying SUSY 
parameters. We find that the distribution of the invariant
mass $m_{ll}$ of the final--state two leptons is sensitive to the
the CP phases as well. First of all, the end point of the maximal invariant
mass is strongly dependent on the CP phases so that after all the
real parameters are determined, the measurement of the end point
will provide us with a very good handle to determine the CP phases.
As one can notice from Figure~15, the sensitivity is larger in the
scenario ${\cal S}1$ with a small $|\mu|$ parameter, which is
more comparable to the value of $M_2$ than in the scenario
${\cal S}2$. It clearly implies that the effect of the CP phases 
is enhanced for comparable gaugino and higgsino parameters.

In passing, we note that since it is independent of the production
mechanism, the lepton invariant mass distribution 
can be an important tool for studying supersymmetric models
even at hadron colliders because of its clean signature.
This point has been in detail explored by Nojiri and Yamada \cite{NY}
by investigating the parameter dependence of the distribution
of the three body decay $\tilde{\chi}^0_2\rightarrow\tilde{\chi}^0_1
\ell^+\ell^-$ at the CERN LHC.

\subsection{Lepton angular distribution in the laboratory frame}

In this section, we give numerical results for the angular distributions
of $\ell^-$ with respect to the electron beam axis computed with
complete spin correlations between production and decay. 
Unlike the invariant mass distribution, this lepton angular distribution
is crucially dependent on the production--decay spin correlations.
Moortgat--Pick and Fraas \cite{MF} have found in their detailed study that
the effect of the spin correlations for the lepton angular distribution
amounts up to 20\% for lower energies and the shape of the lepton 
angular distribution is very sensitive to the mixing in the gaugino
sector and to the value of the slepton mass. These points can be
confirmed by comparing two results displayed in Figure~16. In the 
scenario ${\cal S}1$ (left figure) with large selectron masses,
the lepton angular distribution is forward--backward symmetric and
larger in the forward--backward direction. On the other hand,
in the scenario ${\cal S}2$ the angular distribution is forward--backward 
asymmetric, maximal near $\cos\theta_{\ell^-}=0$ but suppressed in the
forward--backward directions. Furthermore, the size of the 
lepton angular distribution and the forward--backward asymmetry depends 
rather strongly on the CP phases.

\subsection{Triple momentum product}

So far, we concentrate mainly on the CP--even production--decay
correlated observables which depend on the CP phases only indirectly. 
However, the initial electron momentum and two easily--reconstructible
final--state leptons allows us to construct a T--odd observable 
\begin{eqnarray}
{\cal O}_T=\vec{p}_e\cdot\left(\vec{p}_{l^-}\times\vec{p}_{l^+}\right).
\end{eqnarray}
with $l=e,\mu$. This T--observable ${\cal O}_T$ \cite{KOPL} can be 
finite if there exist 
non--vanishing CP--violating or CP--preserving complex phases in the
amplitude for the correlated process. If we neglect the heavy particle 
widths, the T--odd observable can be utilized to directly measure
the CP phases or to constrain them. 

Due to the general property of the T--odd observable, we know that 
it should be given a linear combination of the CP--odd
quartic charges $Q^{21}_4$ and $D_4$. Note that in the scenario ${\cal S}1$
with large slepton masses, both of them are proportional to a small
suppression factor $(s^2_W-1/4)$ as noticed in Eq.~(42). Therefore,
the T--odd observable  as well as the electron EDM can not give any 
significant constraints on the CP phases in the scenario ${\cal S}1$.
On the other hand, since the slepton exchanges give a large contribution
in the scenario ${\cal S}2$, there might be a relatively large value of 
the T-- (CP--)odd observable for no--trivial phases. Furthermore, since the 
fundamental structure of the electron EDM determined by the CP phases will be 
very different from that of the T--odd observable, two CP--odd quantities 
will play a complementary role in constraining the CP phases. In order to 
make a concrete comparison of them, we explore the exclusion region by the 
T--odd observable on the CP phases ${\cal S}2$ in the scenario ${\cal S}2$. 
Since we cannot estimate the precise systematic uncertainties mainly 
related with
the detection quality, we neglect the systematic uncertainties but take into 
account only the statistical errors. In this case, the boundaries of the 
excluded region of the CP phases $\Phi_\mu$ and $\Phi_1$ at the 
$N_\sigma$--$\sigma$ level for a given integrated luminosity satisfy the 
relation
\begin{eqnarray}
\int {\cal L} {\rm d}t =\frac{N^2_\sigma}{2}
   \frac{ \langle {\cal O}^2_T\rangle-\langle {\cal O}_T\rangle^2}{
         |\langle{\cal O}_T\rangle|^2\,\sigma_{tot}}
\end{eqnarray}
where $\langle X\rangle\equiv \int X\frac{{\rm d}\sigma_{tot}}{{\rm d}\Phi}
{\rm d}\Phi/\sigma_{tot}$ over the total phase space volume $\Phi$.
In determining the exclusion area we take into account two possible
combinations of two final--state leptons;$(e^-,e^+)$ and $(\mu^-,\mu^+)$,
which is responsible for the factor 2 in the denominator.

Figure~17 exhibits the excluded area of the CP phases by the electron EDM
measurements at 95\% confidence level (shaded region) and by the T--odd
observable ${\cal O}_T$ (hatched region) at a 2--$\sigma$ level
with an integrated luminosity of 200 fb$^{-1}$. One can clearly notice 
their complementarity role played by two independent CP--odd quantities.
The T--odd observable enables us to exclude all the range of $\Phi_\mu$
for most values of $\Phi_1$ except for $\Phi_1 =0,\pi,2\pi$.
In addition, they are complementary in the sense that the electron EDM is an 
indirect physical quantity determined at a very low energy, which does not 
need to observe SUSY particles but the T--odd observable ${\cal O}_T$
is a direct observable to measure the SUSY CP phases exclusively,
which however requires producing neutralinos directly. 
So, we conclude that through our investigations the T--odd observable
can be a very efficient and complementary quantity in constraining or
determining the CP phases if the lightest and second lightest neutralinos 
are pair--produced and unless the sleptons are too heavy.

\section{Conclusions}

In this paper, we have investigated the associated production of neutralinos  
$e^+e^-\rightarrow\tilde{\chi}^0_1\tilde{\chi}^0_2$ accompanied by the
neutralino leptonic decay $\tilde{\chi}^0_2\rightarrow\tilde{\chi}^0_1
\ell^+\ell^-$, taking into account initial beam polarization 
and production-decay spin correlations in the minimal supersymmetric
standard model with general CP phases but without generational mixing 
in the slepton sector. The stringent constraints by the electron EDM
on the CP phases have been also included in the discussion of the
effects of the CP phases. 

First of all, we have described possible flavor--preserving --
selectron, chargino and neutralino -- mixings in the MSSM with general 
CP phases without generational mixing and applied them to the evaluation 
of the electron EDM to investigate its dependence on the phases.
As a result, we have identified two typical scenarios; one has large
selectron masses of the order of 10 TeV and the other relatively 
light selectron masses of 200 GeV and a large higgisino mass parameter.
The first scenario allows the full range for the CP phases $\Phi_\mu$ and
$\Phi_\mu$ relevant to the neutralino process while the second scenario 
allows a finite space for the CP phases. Employing the allowed space
as the platform for further investigations of the neutralino processes,
we have obtained several interesting results, which can be summarized
as follows:
\begin{itemize}
\item  The production cross section and the branching fractions of the 
       leptonic neutralino decays are very sensitive to the CP phases.
       As a result, the total cross section is very sensitive to the
       CP phases. 
\item  If the electron/positron masses are neglected, the initial longitudinal
       and transverse polarizations  of the initial electron and positron beams
       lead to three CP--even distributions and one CP--odd distribution, 
       which can be studied independently of the details of the neutralino 
       decays. While they are sensitive to the real SUSY 
       parameters, those observables, especially, the T-- (CP--)odd observable
       ${\cal P}_N$ is (almost) insensitive to the CP phases in 
       both scenarios.
\item  The production--decay spin correlations lead to several CP--even 
       observables among them we have studied the two--lepton invariant mass 
       distribution, the lepton angular distribution, and one interesting 
       T--odd (CP--odd) triple product of the initial electron momentum and 
       two final lepton momenta. On the whole, we have found that the 
       distributions are sensitive to the CP phases in the scenario 
       ${\cal S}2$ with relatively light selectrons and large gaugino 
       compositions of the neutralinos. 
\item  We have presented the exclusion region of the CP phases $\Phi_\mu$ and
       $\Phi_1$ by the T--odd (CP--odd) observable with the assumed integrated
       luminosity of 200 fb$^{-1}$ at 2-$\sigma$ level. In comparison with
       the constraints from the electron EDM measurements, the constraints
       from the T--odd observable is complementary in that it constrains very
       strongly the phase $\Phi_1$.
\end{itemize}

To conclude, the associated production of neutralinos $e^+e^-\rightarrow
\tilde{\chi}^0_2\tilde{\chi}^0_1$ followed by the leptonic $\tilde{\chi}^0_2$ 
decays $\tilde{\chi}^0_2\rightarrow\tilde{\chi}^0_1\ell^+\ell^-$ is expected
to be one of the cleanest SUSY processes to allow for a detailed investigation 
of the physics due to the CP phases in the MSSM. Therefore, if 
the neutralinos are produced at future $e^+e^-$ colliders, the colliders will
make it possible to measure or constrain the SUSY parameters and CP phases 
and so provide a complementary check for the existence of CP violation
in the MSSM in the neutralino sector.

\section*{Acknowledgments}

The authors are grateful to Monoranjan Guchait for valuable comments.
S.Y.C. would like to thank Manuel Drees for useful suggestions
and helpful discussions for the present work and also thank Francis 
Halzen and the Physics Department, University of Wisconsin--Madison 
where part of the work has been carried out. The work of H.S.S and
W.Y.S was supported in part by the Korea Science and Engineering 
Foundation (KOSEF) through the KOSEF--DFG large collaboration project, 
Project No.~96--0702--01-01-2, and in part by the Center for 
Theoretical Physics. S.Y.C wishes to acknowledge the financial 
support of 1997-sughak program of Korean Reseach Foundation.



\addtocounter{figure}{1}

\begin{figure}
\begin{center}
\hbox to\textwidth{\hss\epsfig{file=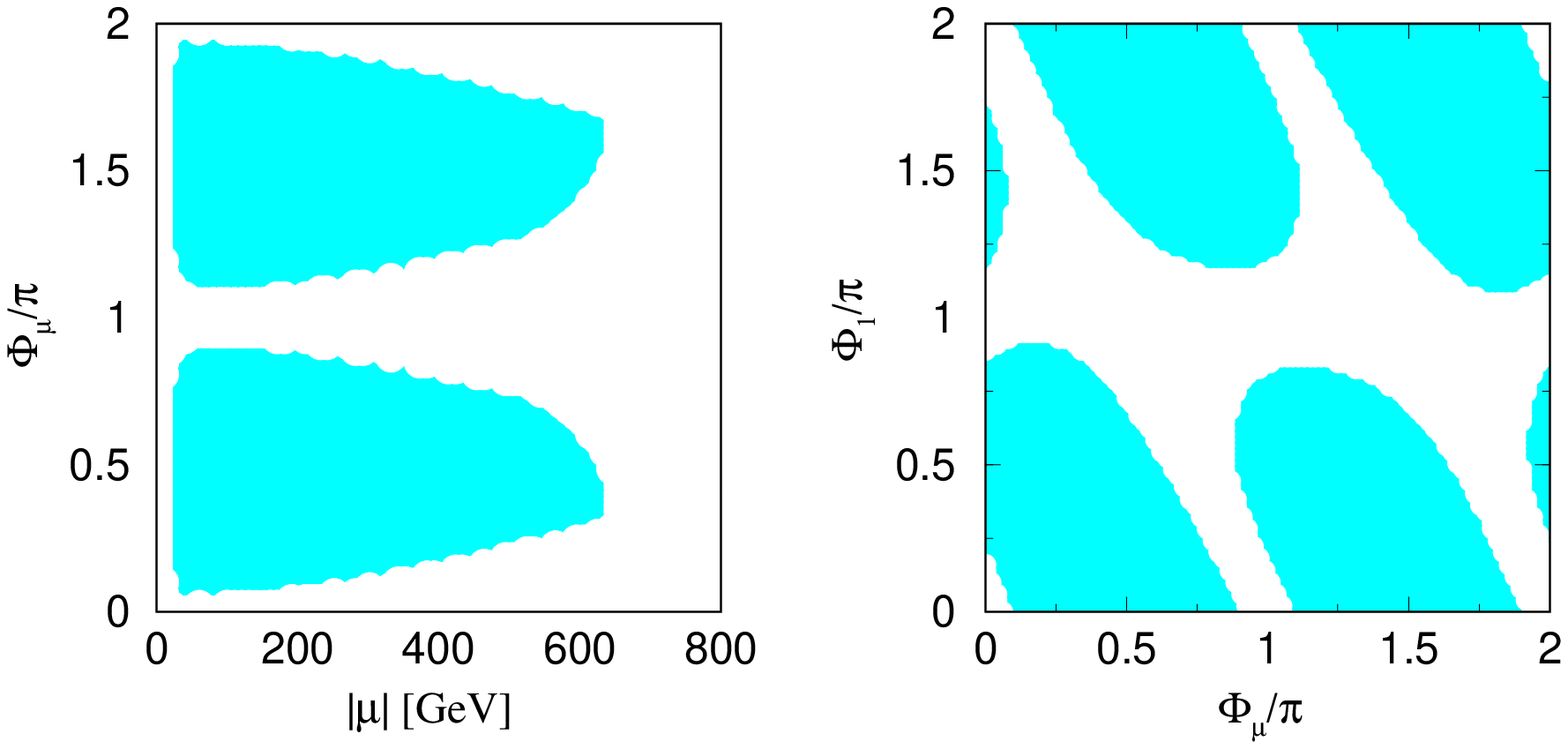,width=16cm,height=7cm}\hss}
\end{center}
\caption{(a) the allowed range for the phase $\Phi_\mu$ vresus the higgsino 
         mass parameter $|\mu|$ and (b) the allowed region of the phases
         $\Phi_\mu$ and $\Phi_1$ against the electron EDM constraints.
         The trilinear parameter $|A_e|$ is taken to be 1 TeV and it phase
         $\Phi_{A_e}$ is scanned over the full allowed range.}
\label{fig2}
\end{figure}

\addtocounter{figure}{1}

\begin{figure}
\begin{center}
\hbox to\textwidth{\hss\epsfig{file=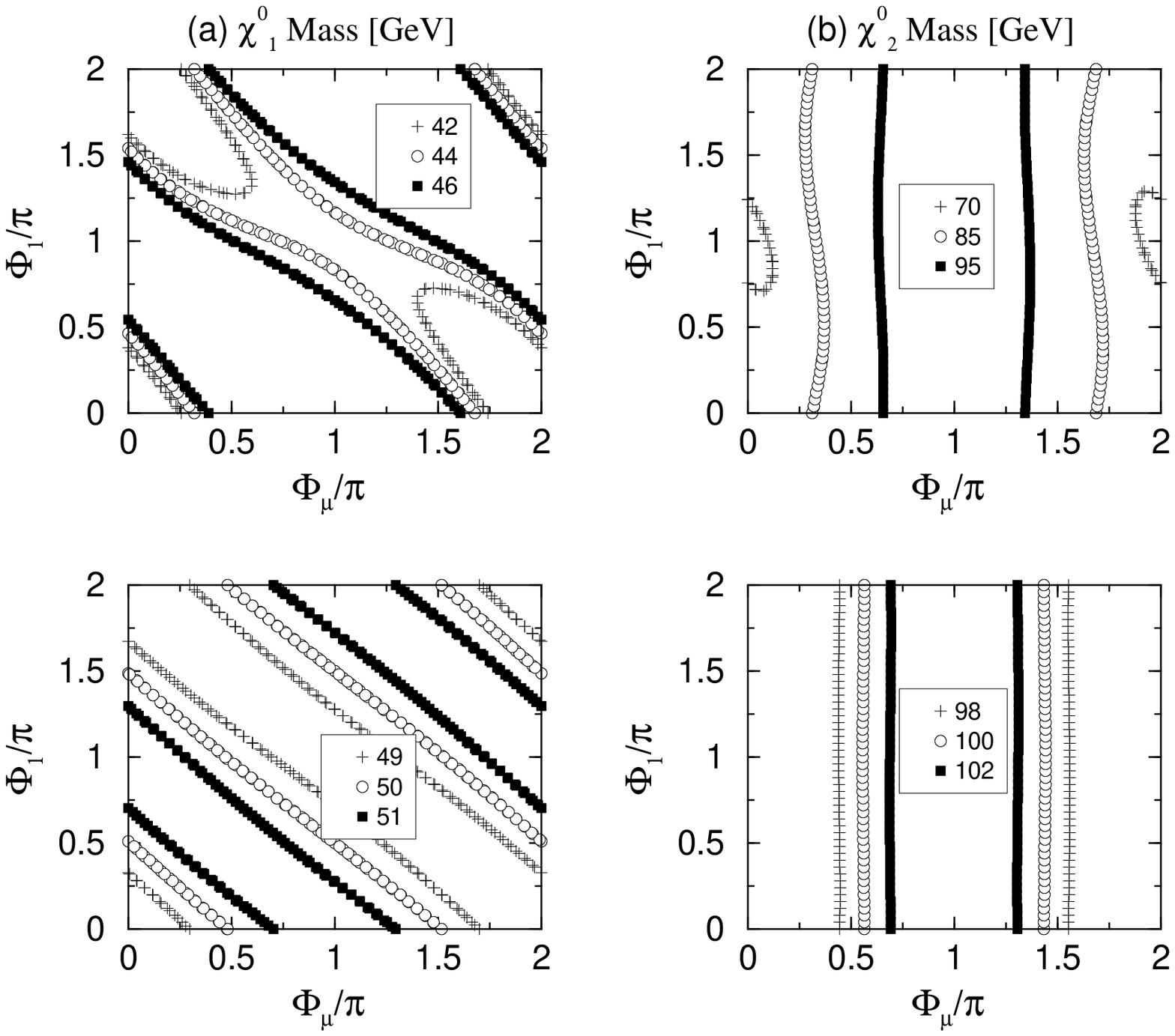,width=12cm,height=10cm}\hss}
\end{center}
\caption{ Neutralino mass spectrum; (a) $m_{\tilde{\chi}^0_1}$ and 
          (b) $m_{\tilde{\chi}^0_2}$ on the $\{\Phi_\mu,\Phi_1\}$ plane 
          in the scenarios ${\cal S}1$ (upper part) and ${\cal S}2$ 
          (lower part).}
\label{fig4}
\end{figure}

\begin{figure}
\begin{center}
\hbox to\textwidth{\hss\epsfig{file=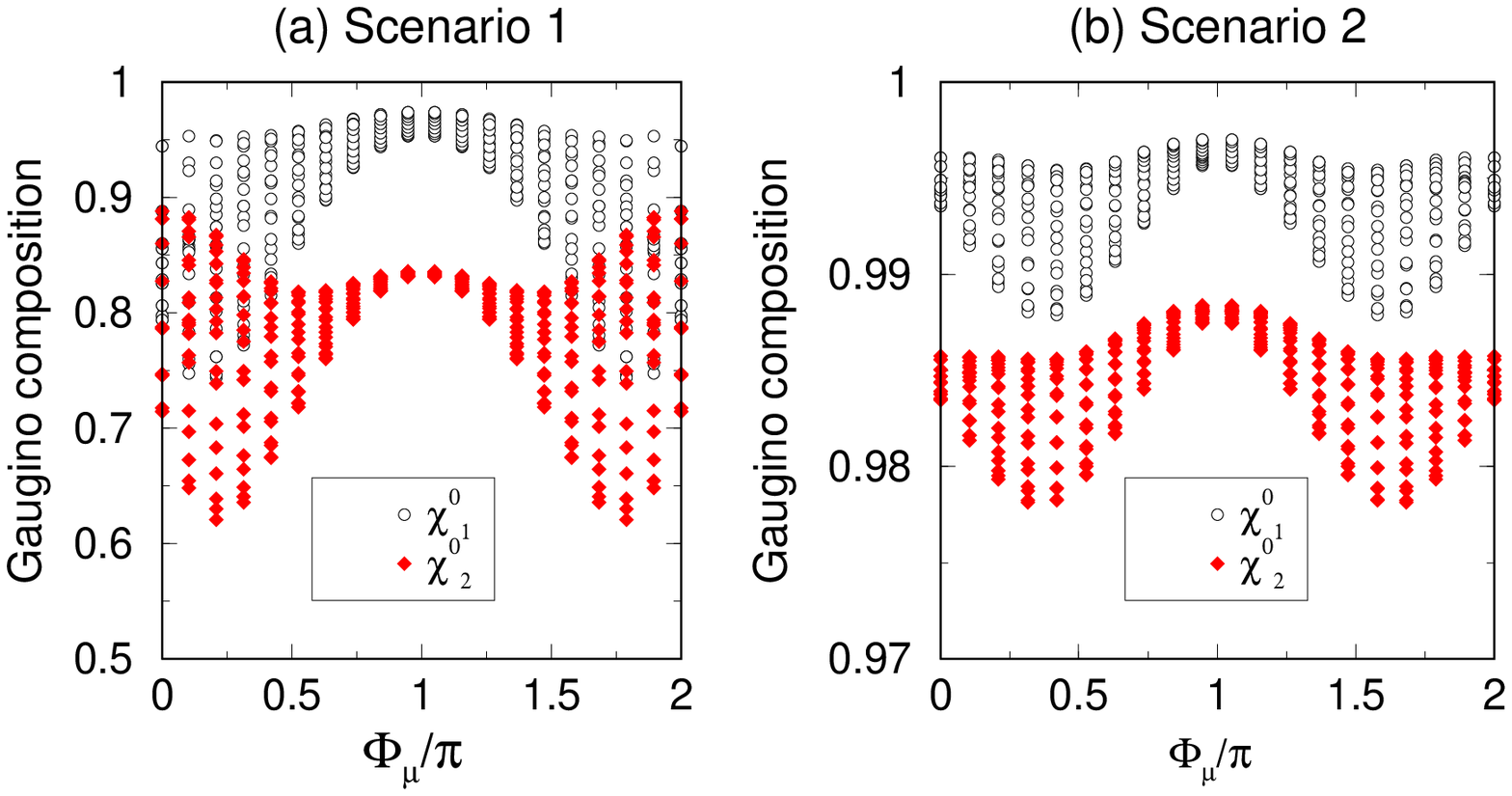,width=14cm,height=7cm}\hss}
\end{center}
\caption{ Gaugino compositions of the two lightest neutralino states 
          $\tilde{\chi}^0_1$ and $\tilde{\chi}^0_2$ with respect to the
          CP phase $\Phi_\mu$ with the phase $\Phi_1$ scanned over its
          full range in (a) in ${\cal S}1$ 
          and (b) ${\cal S}2$. The open circles are for $\tilde{\chi}^0_1$
          and the filled diamonds for $\tilde{\chi}^0_2$.}
\label{fig5}
\end{figure}

\begin{figure}
 \begin{center}
\hbox to\textwidth{\hss\epsfig{file=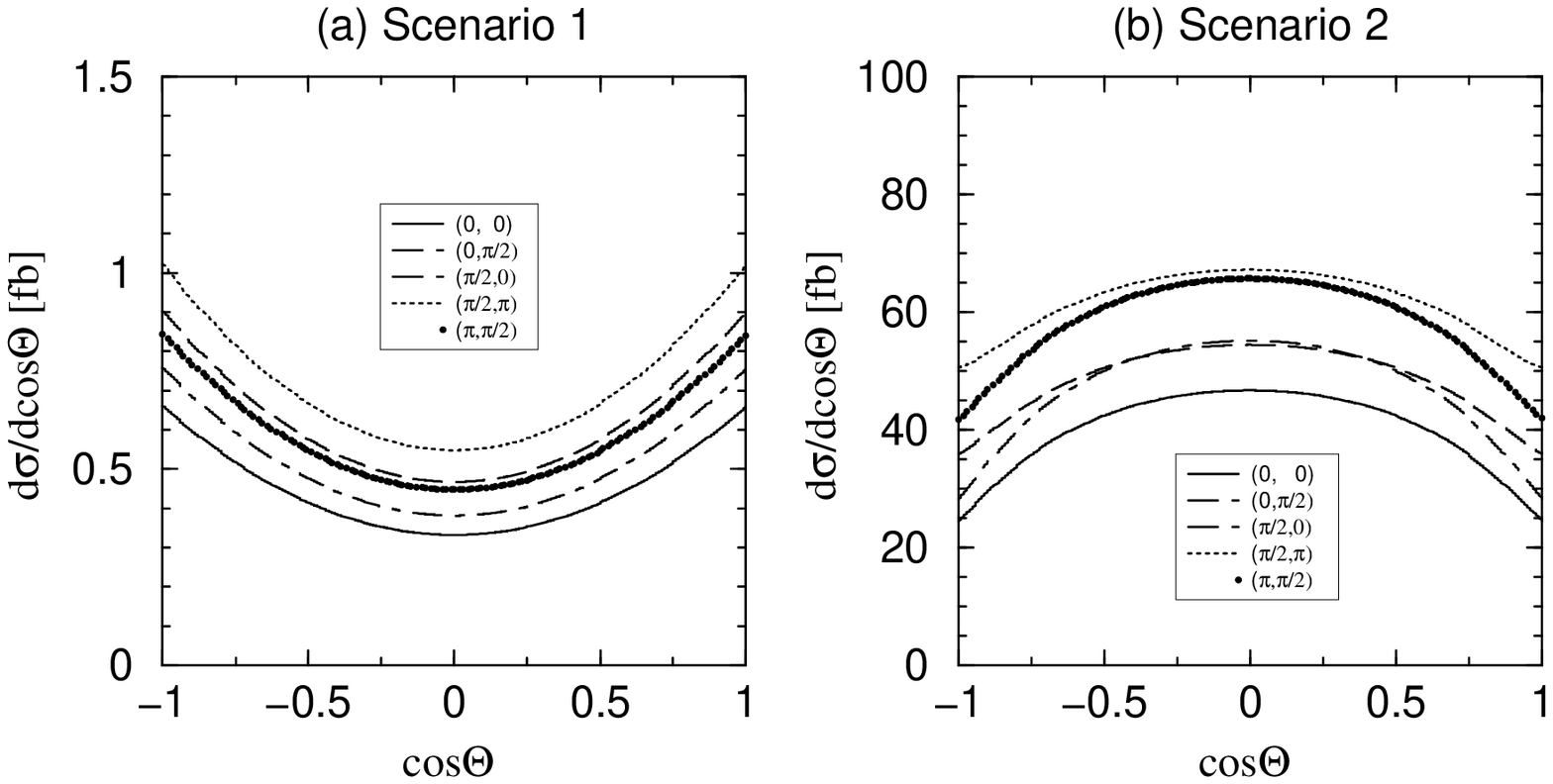,width=14cm,height=7cm}\hss}
 \end{center}
\caption{Dependence of the differential production cross section of the 
         process $e^+e^-\rightarrow\tilde{\chi}^0_2\tilde{\chi}^0_1$ on the 
         scattering angle $\Theta$ in the scenarios (a) ${\cal S}1$ and 
         (b) ${\cal S}2$ for five combinations of the values of two CP phases 
         $\Phi_\mu$ and $\Phi_1$.}
\label{fig6}
\end{figure}

\begin{figure}
\begin{center}
\hbox to\textwidth{\hss\epsfig{file=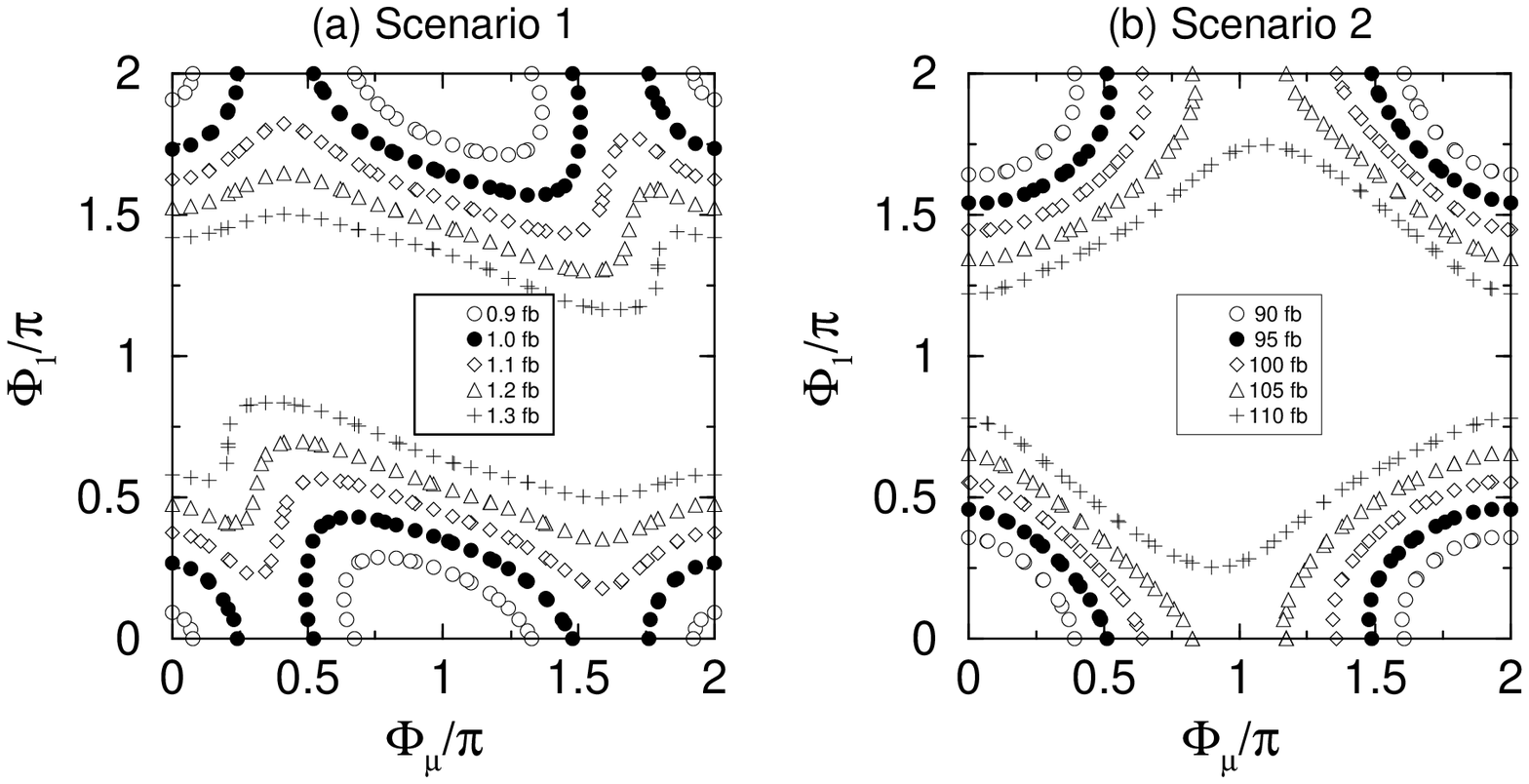,width=15cm,height=7cm}\hss}
\end{center}
\caption{ Dependence of the production cross section $\sigma(e^+e^-\rightarrow
          \tilde{\chi}^0_2\tilde{\chi}^0_1)$ on the $\{\Phi_\mu,\Phi_1\}$ 
          plane in (a) the scenario ${\cal S}1$ and (b) the scenario 
          ${\cal S}2$ for five combinations of the values of two CP 
          phases $\Phi_\mu$ and $\Phi_1$.}
\label{fig7}
\end{figure}

\addtocounter{figure}{1}

\begin{figure}
 \begin{center}
\hbox to\textwidth{\hss\epsfig{file=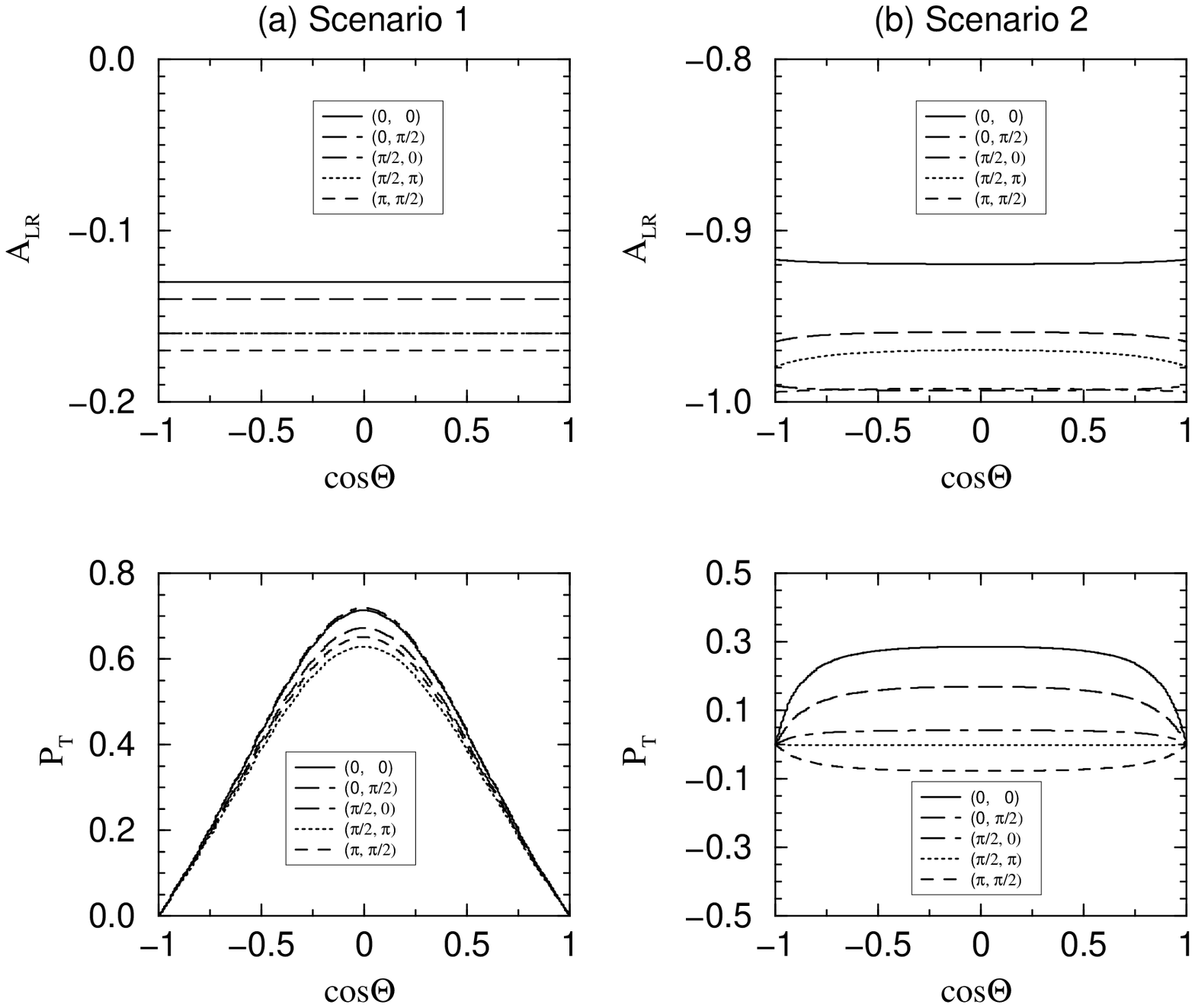,width=10cm,height=10cm}\hss}
 \end{center}
\caption{Dependence of the left--right asymmetry $A_{LR}$ and the transverse
         distribution $P_T$ on the scattering angle  $\Theta$ in (a) the 
         scenario ${\cal S}1$ and (b) ${\cal S}2$ for five combinations of
         the values of two CP phases $\Phi_\mu$ and $\Phi_1$. The upper 
         two figures are for $A_{LR}$ and the lower two figures for $P_T$.}
\label{fig9}
\end{figure}

\begin{figure}
 \begin{center}
\hbox to\textwidth{\hss\epsfig{file=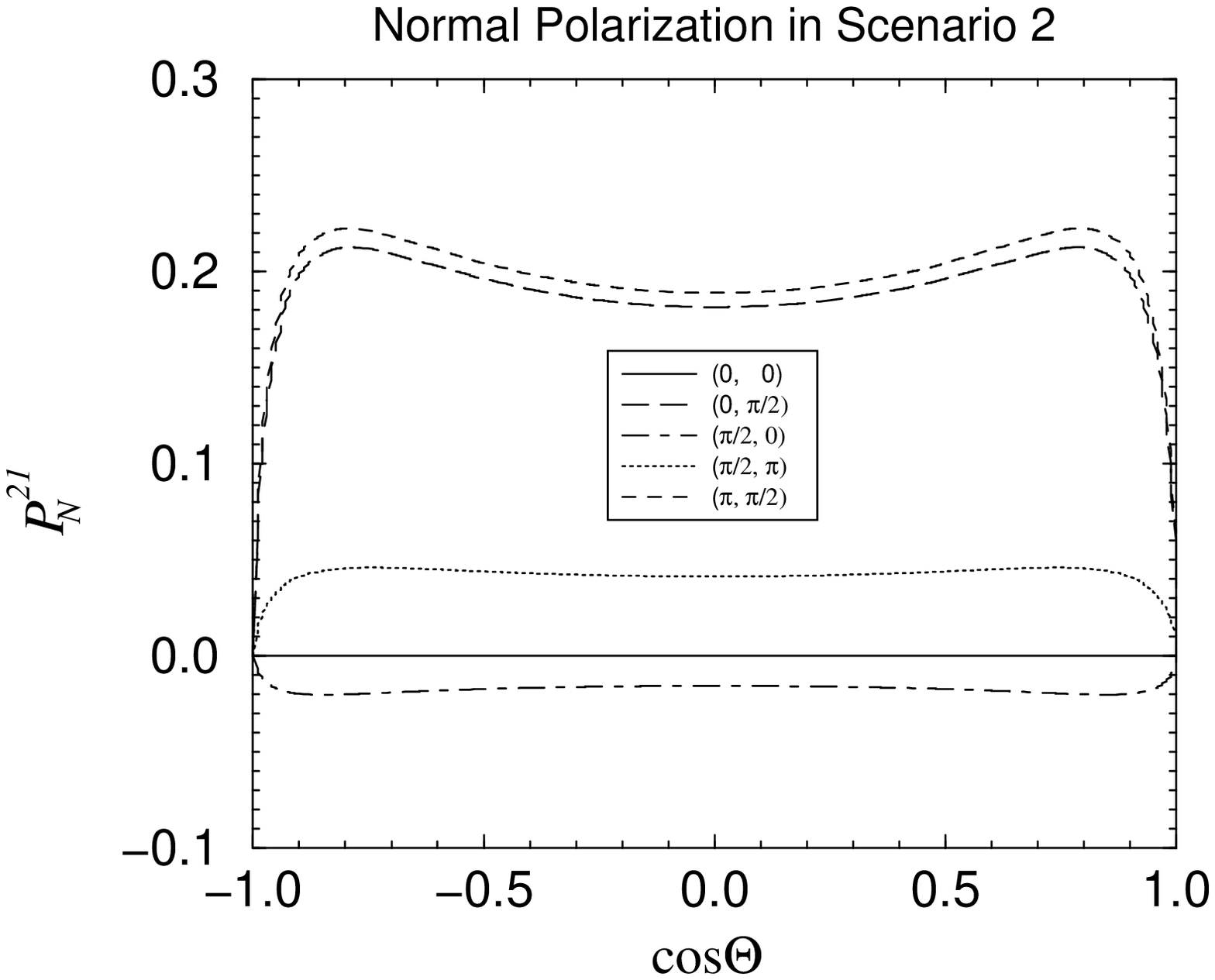,width=8cm,height=8cm}\hss}
 \end{center}
\caption{Dependence of the $\tilde{\chi}^0_2$ normal polarization 
         ${\cal P}^{21}_N$ on the scattering angle $\Theta$ in the scenario 
         ${\cal S}2$ with relatively light selectron masses and a large
         value of $|\mu|=700$ GeV for five combinations of the values of the
         CP phases $\Phi_\mu$ and $\Phi_1$.}
\label{fig10}
\end{figure}

\addtocounter{figure}{1}

\begin{figure}
\begin{center}
\hbox to\textwidth{\hss\epsfig{file=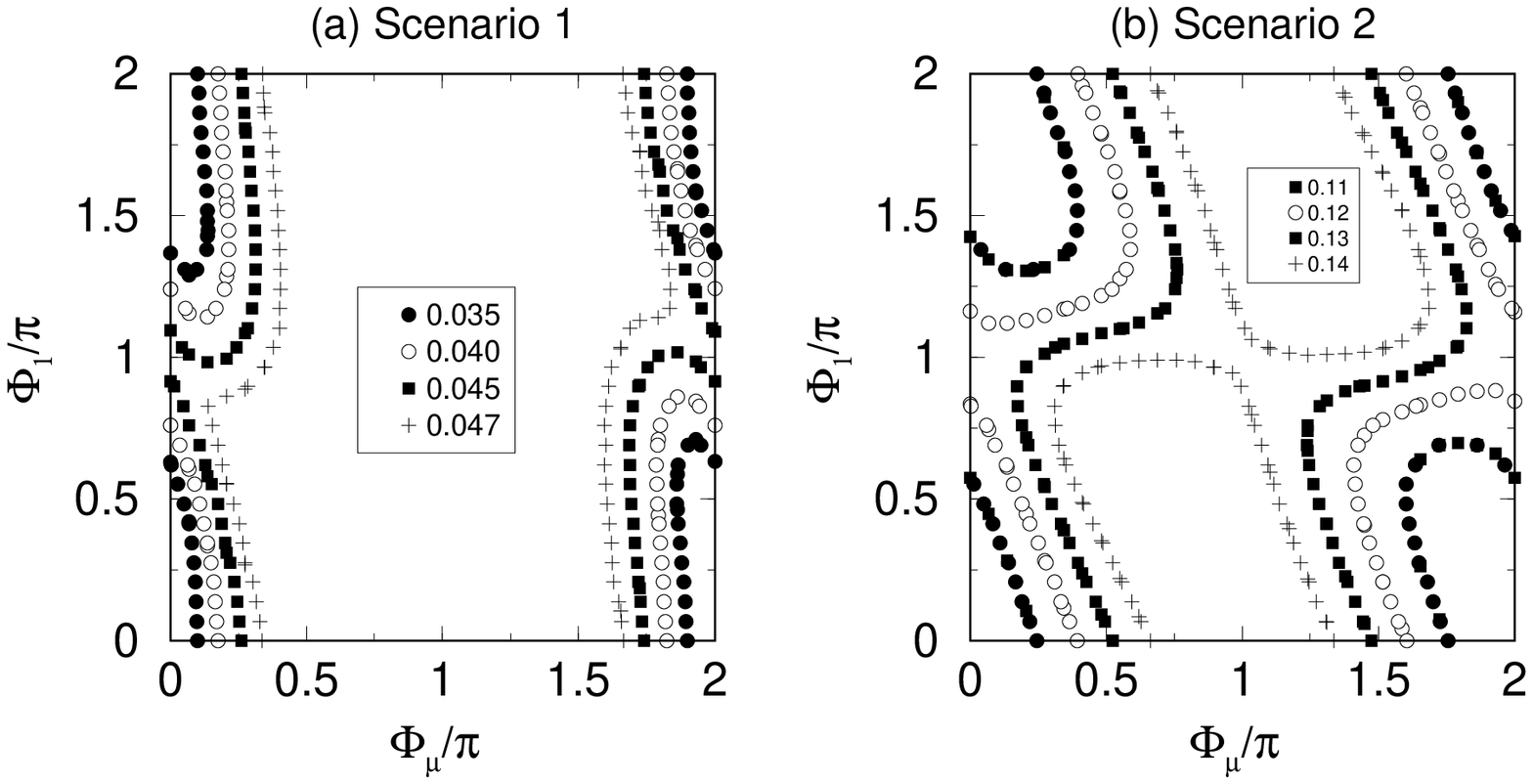,width=15cm,height=7cm}\hss}
\end{center}
\caption{ Contours of the branching fraction ${\cal B}(\tilde{\chi}^0_2
          \rightarrow\tilde{\chi}^0_1\ell^+\ell^-)$ for $\ell=e$ or $\mu$ on
          the $\{\Phi_\mu,\Phi_1\}$ plane in (a) the scenarios (a) ${\cal S}1$
          and (b) the scenario ${\cal S}2$.}
\label{fig12}
\end{figure}

\addtocounter{figure}{1}

\begin{figure}
\begin{center}
\hbox to\textwidth{\hss\epsfig{file=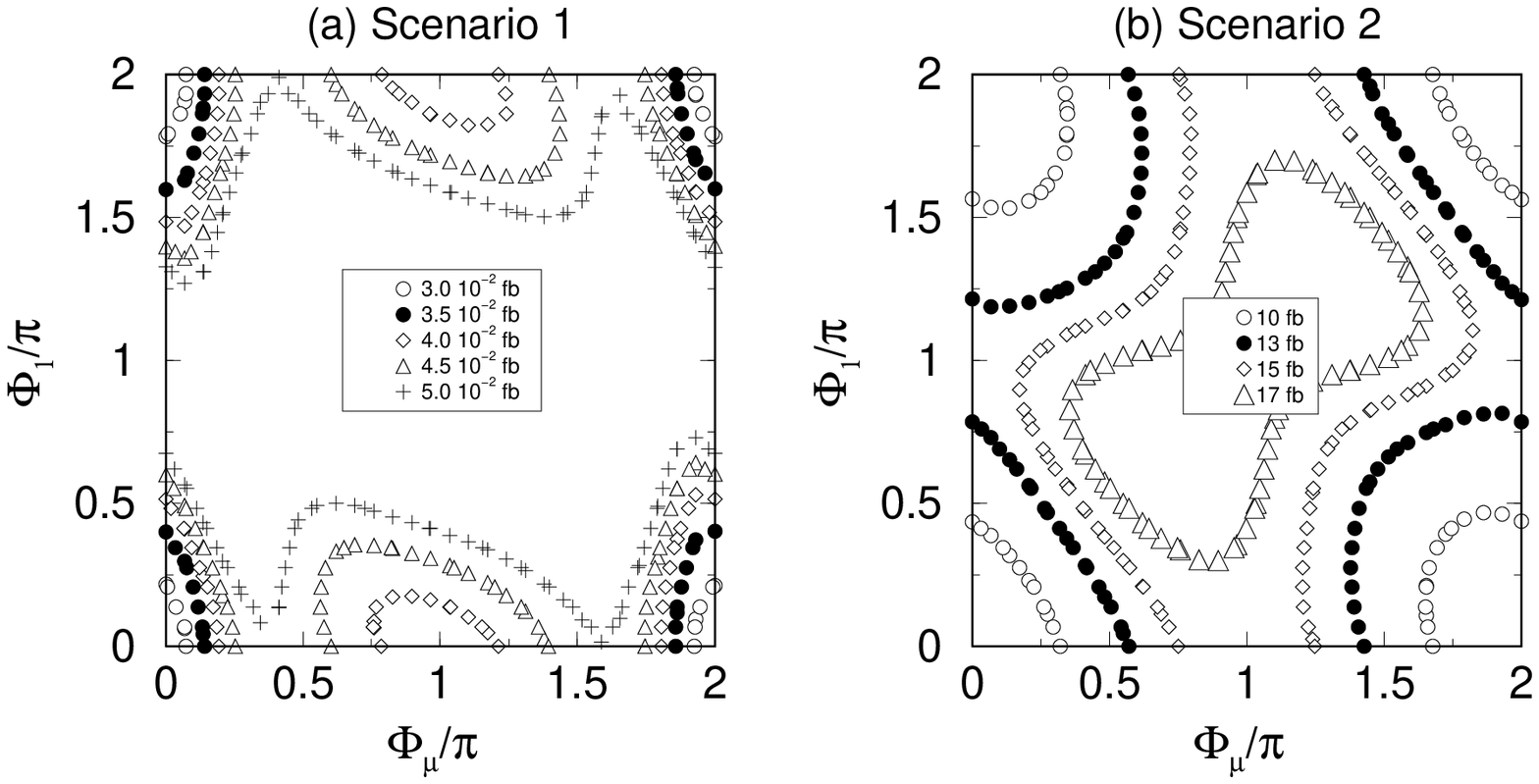,width=16cm,height=7cm}\hss}
\end{center}
\caption{ Contours of the total cross section of the associated production
          of neutralinos $e^+e^-\rightarrow\tilde{\chi}^0_2
          \tilde{\chi}^0_1$ followed by the sub-sequential decay
          $\tilde{\chi}^0_2\rightarrow\tilde{\chi}^0_1\ell^+\ell^-$ 
          on the plane of two phases $\Phi_\mu$ and $\Phi_1$ in 
          the scenarios (a) ${\cal S}1$ and (b) ${\cal S}2$. Here,
          $\ell$ is either $e$ or $\mu$.  }
\label{fig14}
\end{figure}

\begin{figure}
\begin{center}
\hbox to\textwidth{\hss\epsfig{file=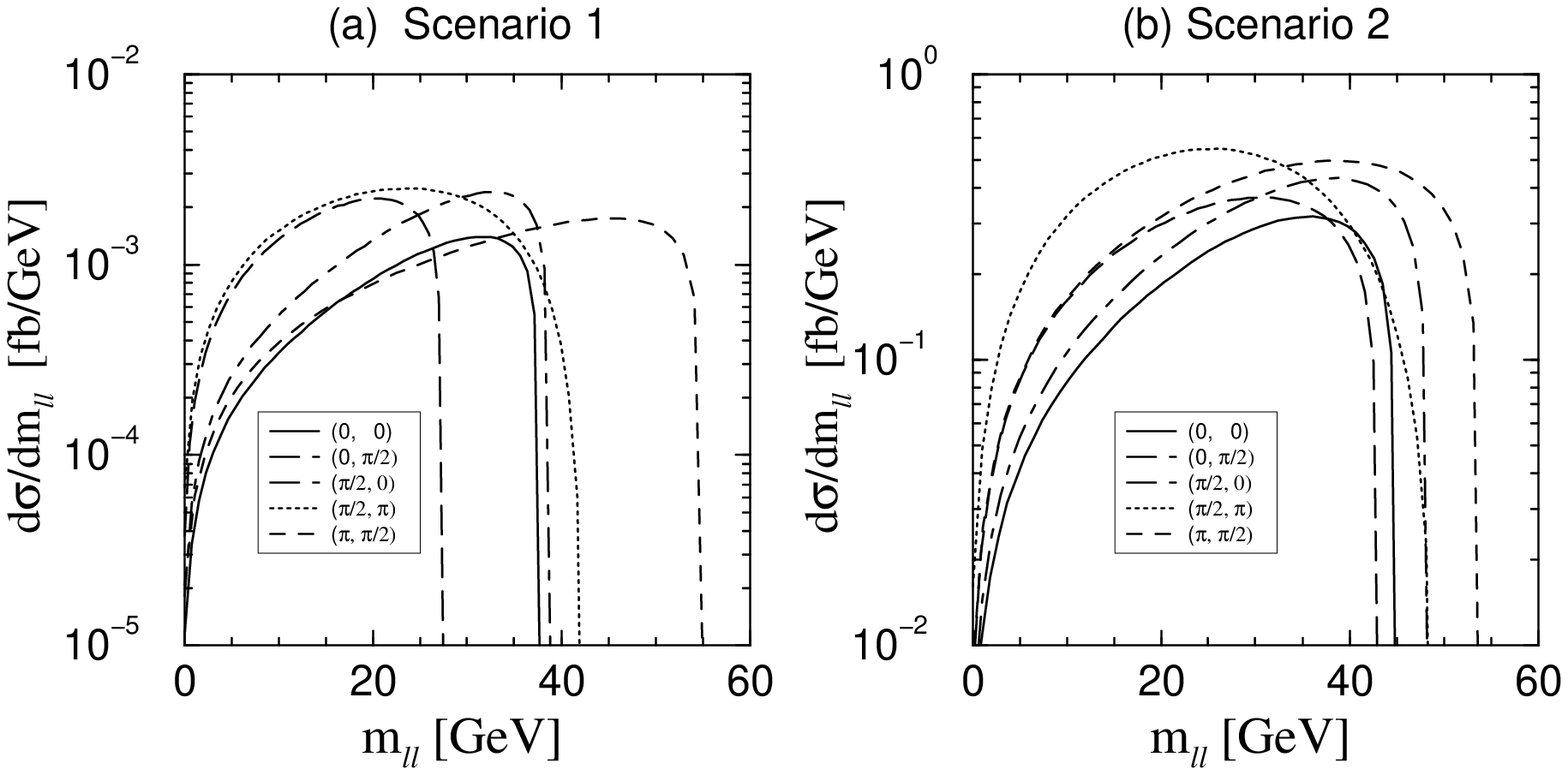,width=16cm,height=7cm}\hss}
\end{center}
\caption{ Two--lepton invariant mass distributions of the correlated production
          process $e^+e^-\rightarrow\tilde{\chi}^0_1\tilde{\chi}^0_2\rightarrow
          \tilde{\chi}^0_1\ell^+\ell^-$ in the scenarios (a) ${\cal S}1$ and 
          (b) ${\cal S}2$ for five combinations of the values of two
           CP phases $\Phi_\mu$ and $\Phi_1$. }
\label{fig15}
\end{figure}

\begin{figure}
\begin{center}
\hbox to\textwidth{\hss\epsfig{file=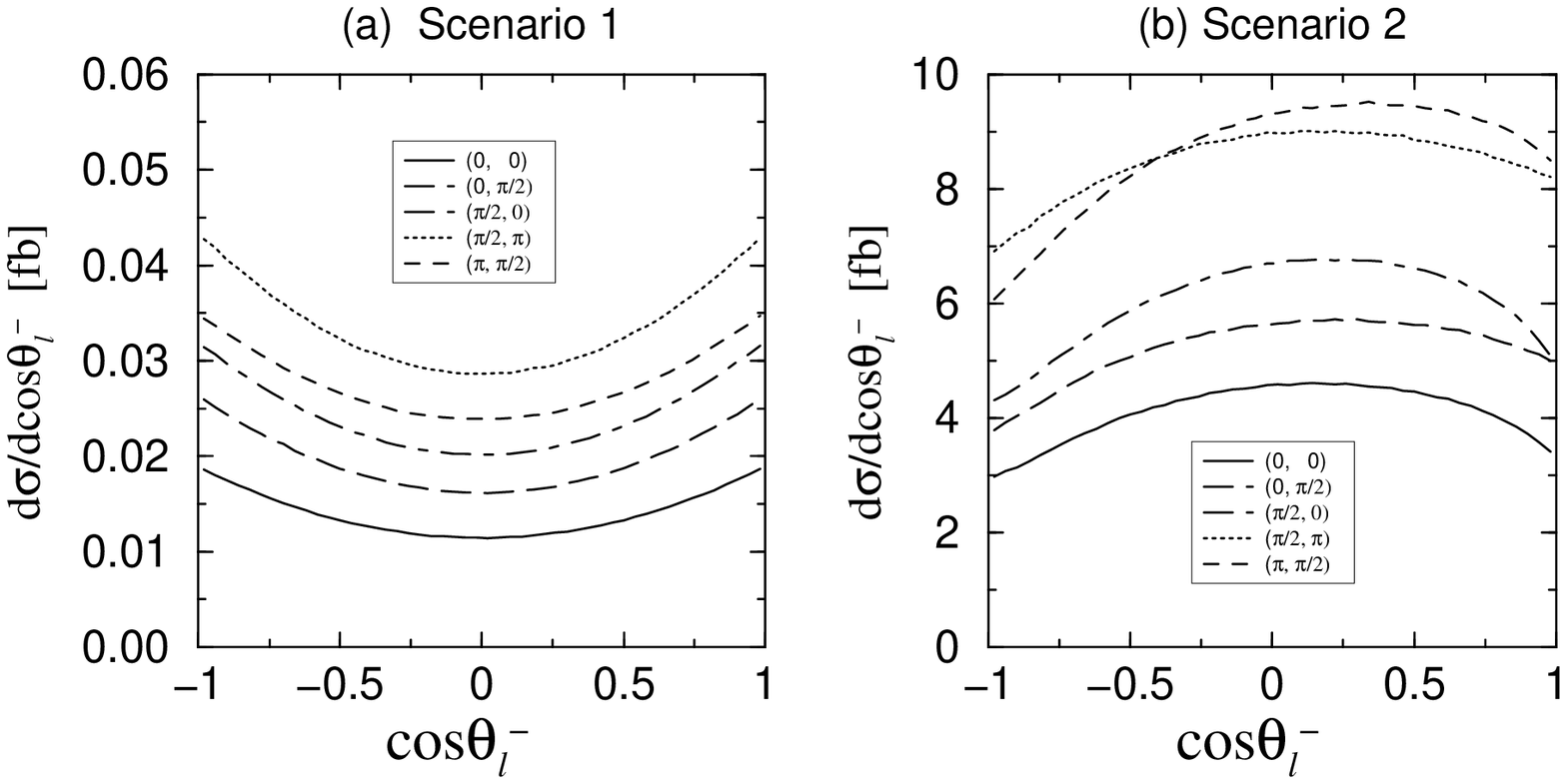,width=16cm,height=7cm}\hss}
\end{center}
\caption{ Lepton angular distributions of the correlated process 
          $e^+e^-\rightarrow\tilde{\chi}^0_1\tilde{\chi}^0_2\rightarrow
          \tilde{\chi}^0_1(\tilde{\chi}^0_1\ell^+\ell^-)$ in the scenarios 
	  (a) ${\cal S}1$ and (b) ${\cal S}2$ for five combinations of the 
	  values of two CP phases $\Phi_\mu$ and $\Phi_1$.}
\label{fig16}
\end{figure}

\begin{figure}
\begin{center}
\hbox to\textwidth{\hss\epsfig{file=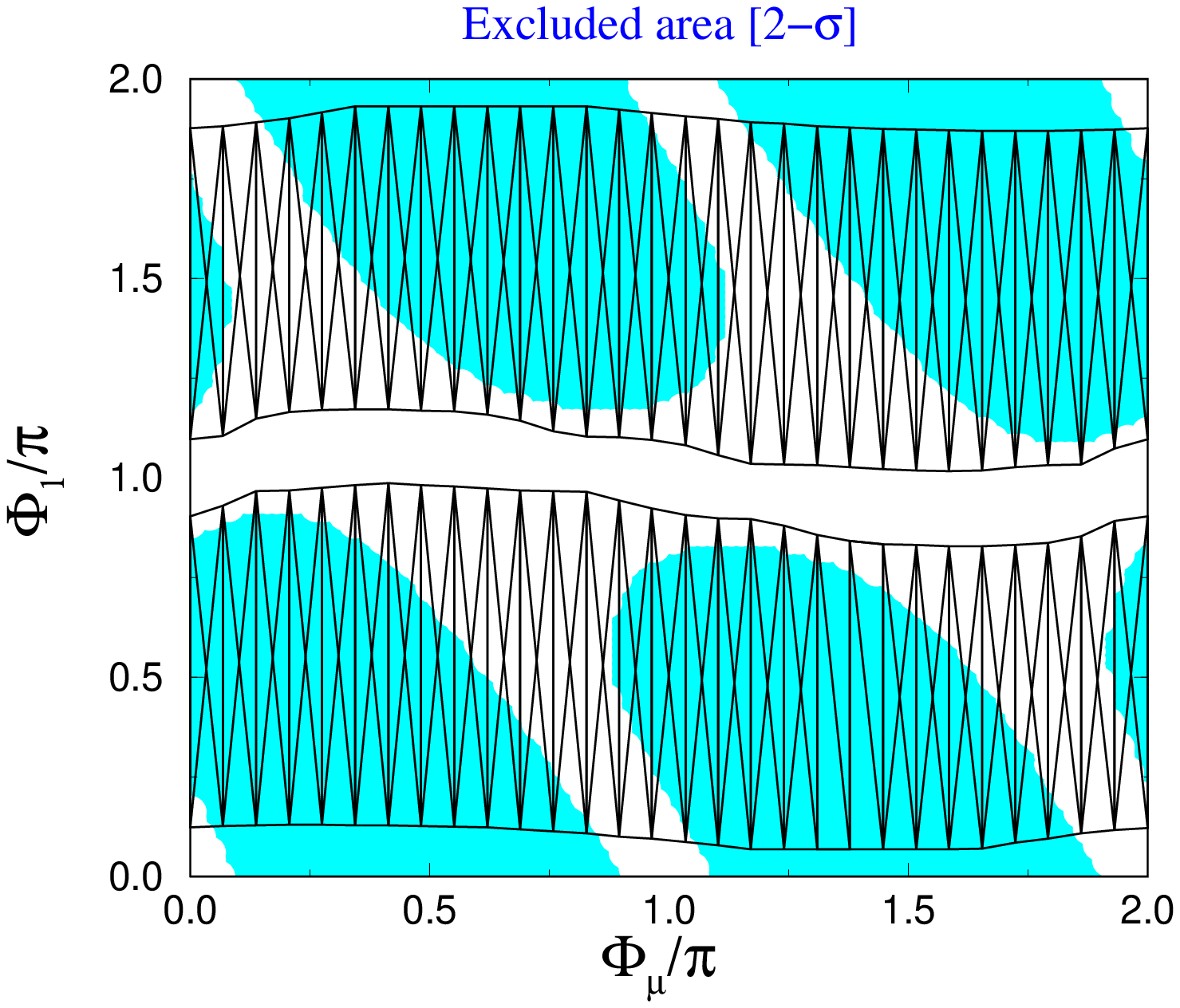,width=8cm,height=8cm}\hss}
\end{center}
\caption{ Excluded region of the phases $\{\Phi_\mu,\Phi_1\}$ by the electron 
          EDM constraints (shadowed region) and by the triple product 
          measurements with the integrated luminosity of 200 fb$^{-1}$
          at the 2--$\sigma$ level.}
\label{fig17}
\end{figure}

\vfil\eject

\end{document}